\documentclass[12pt]{article}
\usepackage{psfig}
\begin{document}
\title{Strangeness Production in the HSD Transport Approach
from SIS to SPS energies\footnote{Supported 
by BMBF and GSI Darmstadt}}
\author{J.~Geiss\footnote{Part of the 
PhD thesis of J. Geiss}, W. Cassing and C. Greiner \\ 
Institut f\"ur Theoretische Physik, Universit\"at Giessen \\
D-35392 Giessen, Germany}
\date{ }
\maketitle

\begin{abstract}
We study systematically the production of strangeness in nuclear reactions from
SIS to SPS energies within the covariant hadronic transport approach HSD. 
Whereas the proton and pion rapidity distributions as well as pion
transverse momentum spectra are well described in the hadronic transport
model from 2-200 A$\cdot$GeV, the $K^+$ and $K^-$ spectra are noticeably 
underestimated at
AGS energies while the $K^+$ spectra match well at SIS and SPS 
energies with the experimental 
data. We conclude that the failure of the hadronic model at AGS energies 
points towards a nonhadronic phase during the collision of heavy
systems around 10 A$\cdot$GeV.

\end{abstract}

\noindent

\section{Introduction}
The aim of high energy heavy-ion collisions especially at the Brookhaven 
Alternating Gradient Synchrotron (AGS) and the CERN Super Proton 
Sychrotron (SPS) is to investigate nuclear matter under extreme
conditions, i.e. high temperature and high density. The most exciting
prospect is the possible observation of a signal for a phase 
transition from normal nuclear matter to a nonhadronic phase, where
partons are the basic degrees of freedom. In this context Rafelski
has introduced strangeness enhancement in heavy-ion collisions
compared to proton-proton collisions as a possible signature for
the phase transition \cite{Rafelski1}. Especially the abundancies
of multistrange baryons and strange antibaryons should be increased 
drastically. The idea is based on a 
different production mechanism of strangeness in a partonic   
and hadronic phase, respectively. In the partonic phase 
the mass of the strange 
quark is of the order of the temperature resulting in a rapid 
chemical equilibration of the flavors, $u$ $d$ and $s$, if the partonic phase 
is of sufficient  size and duration to form an equilibrated quark-gluon
plasma (QGP). The timescale for chemical equilibration in a QGP
is estimated to be of the order $2-3$ fm/c 
\cite{Rafelski2,Rafelski3,Tamas}. In contrast,
the equilibration time of strange hadrons in a pure thermally 
equilibrated hadronic phase is expected to be an order of magnitude 
larger \cite{Rafelski2}, because the threshold
for strange hadron production is large compared to typical
temperatures expected in a heavy-ion collision at AGS or SPS energies. 

Indeed it has been found experimentally, 
that the strange hadron yields normalized
to the pion abundancies are enhanced in A~+~A collisions at AGS and SPS energies
\cite{Strangeness}. In particular a strong increase 
of the production of strange antibaryons and multistrange 
baryon/antibaryons has been observed  experimentally. 
Nevertheless, there is still a theoretical debate, if 
the observed strangeness enhancement is a clear signature for a 
QGP. Detailed calculations of the
production mechanism are difficult due the nonperturbative
nature of the hadronization process and phenomenological
models must be introduced to describe hadron condensation.  

Several microscopic approaches have been used to describe 
ultrarelativistic heavy-ion collisions. Most of them describe
the early prehadronic phase and hadronization in terms of the 
string picture for the high energy hadronic interactions
such as the HSD approach \cite{cassing}. Other popular
models of this type are FRITIOF \cite{LUND} or the
recently extended version LUCIAE \cite{LUCIAE}, VENUS \cite{VENUS},
RQMD \cite{RQMD}, DPM \cite{DPM}, QGSM \cite{QGSM}, ARC \cite{ARC},
ART \cite{LiPRC} and UrQMD \cite{Bass}. 
The strings, representing a prehadronic stage, are
characterized by the constituent incomming quarks and a 
tube of color flux is supposed to be spanned in between. 
While in HSD, FRITIOF and UrQMD the strings are assumed to 
hadronize independently, recent
RQMD versions and VENUS versions include some kind of string fusion 
resulting in color ropes \cite{Hsorge,Werner2} 
or quark droplets \cite{Werner3}, which  may be seen as a
'mini-QGP', to achieve a better agreement with
data, especially the strange baryon-antibaryon yield, at SPS energies.
In the RQMD model, for example, the color rope then fragments in 
a collective way and tends to enhance the production of
strange quark pairs and especially of diquark-antidiquark pairs 
\cite{Hsorge2}. The 
production of strange baryon-antibaryon pairs is enhanced drastically 
by this collective effect while the enhancement of the 
total strangeness yield, mainly kaons, at
SPS energies was found to be dominated by hadronic rescattering 
and not by an increasing $s\bar{s}$ yield at the 
hadronization process. In a similar way Tai An and Sa Ben-Hao describe
the enhancement of strange antibaryons and multistrange baryons 
at SPS energies within LUCIAE, which is based on the FRITIOF model 
and includes hadronic rescattering and some collective string 
interaction \cite{LUCIAE}. This collective interaction tends to enhance 
the probability for strangeness and diquark production  
during the string fragmentation process
as a function of centrality and mass of the system
\cite{LUCIAE}. In the present work the strangeness production
in proton-nucleus and heavy-ion collisions is investigated within a dynamical 
hadronic description, the HSD approach \cite{cassing}. No
enhancement of the primary strangeness production during
the hadronization phase is included as well as no kind of string-string
interaction.

Another class of models are the parton models HIJING \cite{HIJING},
HIJET \cite{HIJET}
and VNI \cite{VNI}, which are based on the concept that the colliding
nuclei can be decomposed into their parton substructure. The parton
models include cross sections from perturbative QCD requireing
high momentum transfers $ q^2 > 10$ GeV$^2$ in parton scatterings. At SPS
energies this may be fullfilled for the first collisions, but most
of the reactions are supposed to be softer and require a nonperturbative
treatment. For that reason the application of parton models at 
SPS energies remains questionable; the situation will become
better at RHIC and LHC energies. 

In contrast to microscopic models also (global) thermal models 
have been employed due to their simplicity 
\cite{BraunMu,Letessier,Sollfrank,greiner}. Here the
hot and dense system is assumed to be in local thermal
and chemical equilibrium at the time of freeze-out. 
Thermal models state nothing about the history of the fireball,
if it was created by a pure hadronic system or a QGP.
Thus there is a total 'memory loss'. Nevertheless,
thermal models are attractive, because they predict all
particle abundancies in terms of only three parameters, the 
temperature T, the baryochemical potential $\mu_B$ and the volume of 
the fireball $V_f$. Absolute particle number ratios 
reflect the thermal and chemical conditions at freezout, while the 
spectra of particles combine thermal motion
and collective flow. A lot of data 
at AGS and SPS energies has been found to be consistent
with  thermal models within a factor of 2
\cite{BraunMu}. For the strangeness production a new incomplete
strangeness saturation parameter  has to be included \cite{Letessier}
in order to refine the agreement. 

On the other hand, many of the dynamical microscopic models are not 
designed to work over a broad energy range.
The VENUS model is fine tuned to SPS energies and ARC is presently only 
designed for AGS energies. An extention to higher energies 
including hard processes as described in \cite{Kahana} will increase
the applicability of ARC in future. 
The parton models on the other hand are restricted
to very high energies as mentioned above. On the other hand 
a systematic study over a wide energy range (SIS-SPS) 
from light to heavy systems is of particular interest 
for the topic of strangeness enhancement in heavy-ion collisions,
because recent data from AGS and SPS show that the scaled strangeness
yield ($K/\pi$) seems to be essentially higher at AGS energies.
Such a systematic investigation can presently only be performed within the 
HSD, RQMD or UrQMD models. Within the RQMD model there exist
calculations for strangeness production at AGS \cite{Gonin} and
at SPS energies \cite{Hsorge,Hsorge2,Sorge3}, but these investigations are 
performed within different RQMD versions. 

The aim of the present work is to perform  a systematic analysis
of strangeness production within the HSD approach and to present
the elementary strangeness production channels. Recently a lot
of data for the most heavy systems has been published \cite{Strangeness};
thus it is possible to compare our results to the data for many 
systems from SIS to SPS energies within one fixed version of the 
transport code.

Our work is organized as follows: In Section 2 we will give a brief 
description of the HSD approach. The primary elementary 
hadron-hadron collisions are described in detail and the dynamics
of the strings in heavy-ion collisions as well as the
formation time concept is presented. 
The particle production and stopping of incoming hadrons
for elementary baryon-baryon collisions is investigated over a wide energy 
range. We then extend our study to the strangeness production  
from string fragmentation and from rescattering processes. We
compare our results with data from proton-proton and pion-proton
collisions from the threshold up to 100 GeV. In Section 3 
we test the global dynamics from AGS to SPS energies obtained
in the HSD model for  different systems. 
We compare our results for stopping and 
particle production to data and show  that
the global dynamics is reasonably well described within the
HSD approach, which also was found  at SIS energies 
before \cite{cassing2,cassing3}. In Section 4 we show our 
results for strangeness production
in heavy-ion collisions at AGS and SPS energies in comparison with 
the available data.
In Section 5 we present the results of our calculation for pion and kaon
spectra from 2 to 11 A$\cdot$GeV thus providing a link between SIS
and AGS energies. 
In Section 6, finally, we give a summary and discussion of open problems.

\section{The covariant transport approach}
In this work the dynamical analysis of p~+~p, p~+~A and A~+~A reactions
is performed within the HSD approach 
\cite{cassing} in the cascade modus 
which is based on a coupled set of covariant transport
equations for the phase-space distributions $f_{h} (x,p)$ of hadron $h$
\cite{cassing}, i.e.
\begin{samepage}
\begin{eqnarray}  \label{g24}
\lefteqn{\left( {\partial \over \partial t} + {\vec{p}_1 \over m} 
\vec{\nabla} \right)
f_1(x,p_1)  } \nonumber \\
&&= \sum_{2, 3, 4\ldots}\int d2 d3 d4 \ldots
 [G^{\dagger}G]_{12\to 34\ldots}
\delta^4(p_1^{\mu} +p_2^{\mu} -p_3^{\mu} -p_4^{\mu}\ldots )  \nonumber\\
&& \times \left\{ f_3(x,p_3)f_4(x,p_4)\bar{f}_1(x,p)
\bar{f}_2(x,p_2)\right.  \nonumber\\
&& -\left. f_1(x,p)f_2(x,p_2)\bar{f}_3(x,p_3)
\bar{f}_4(x,p_4) \right\} \ldots\ \ .
\end{eqnarray}
\end{samepage}
Here $ [G^{\dagger}G]_{12\to 34\ldots}
\delta^4(p_1^{\mu} +p_2^{\mu} -p_3^{\mu} -p_4^{\mu}\ldots )$ 
is the `transition rate' for the process
$1+2\to 3+4+\ldots$,
while the phase-space factors
\begin{equation}
\bar{f}_{h} (x,p)=1 \pm f_{{h}} (x,p)
\end{equation}
are responsible for fermion Pauli-blocking or Bose enhancement,
respectively, depending on the type of hadron in the final/initial
channel. The dots in eq.~(\ref{g24}) stand for further contributions in
the collision term with more than two hadrons in the final
channels. We note that collisions with more than two hadrons in the 
initial channel are not included due to technical reasons.

The transport approach (\ref{g24}) is fully specified 
by the transition rates $G^\dagger G\,\delta^4 (\ldots )$ in the collision
term, that describes the scattering and hadron production and absorption
rates. In the cms of the colliding particle the transition rate is given by
\begin{eqnarray}
G^\dagger G\,\delta^4 (\ldots )=v_{12} \left. {d\sigma \over d \Omega}
\right|_{1+2\to 3+4+\ldots},
\end{eqnarray}
where $d\sigma /d \Omega$ is the differential cross section of the
reaction and $v_{12}$ the relative velocity of particles 1 and 2. 
The HSD transport approach
was found to describe reasonably well hadronic as well as dilepton
data from SIS to SPS energies \cite{cassing,cassing2,cassing3,cassing1}.

In the present version we propagate explicitly baryons
(p, n, $\Delta$, N(1440), N(1535), $\Lambda$, $\Sigma$, $\Sigma^*$, $\Xi$,
$\Omega$), the corresponding antibaryons and
mesons (pions, kaons, $\eta$'s, $\eta^\prime$'s, $\rho, \omega, \phi$, 
$K^*$, $a_1$). The baryon-baryon collisions are described using the explicit
cross sections as in the BUU model \cite{Wolf} for invariant energies 
$\sqrt{s} < 2.65$ GeV, which have been successfully tested in the energy 
regime below 2 A$\cdot$GeV bombarding energy - and by the 
FRITIOF model \cite{LUND} (for $\sqrt{s} > 2.65$ GeV). 
For meson-baryon reactions the same concept is used,
where a transition energy of $2.1$ GeV is employed. In the BUU model
\cite{Wolf} the following reaction channels are included
\begin{eqnarray}
NN&\leftrightarrow&N'N' \\ \nonumber
NN&\leftrightarrow&NR \\ \nonumber
NR&\leftrightarrow&N'R' \\ \nonumber
\Delta \Delta &\leftrightarrow&N'R' \\ \nonumber
R &\leftrightarrow& N\pi \\ \nonumber
N(1535) &\leftrightarrow& N\eta \\ \nonumber
NN&\leftrightarrow&NN \pi,  
\end{eqnarray}
where $R$ stands for a resonance $\Delta, N(1440)$ or $N(1535)$. 
Additionally the channels 
\begin{eqnarray}
\rho N &\to & N\pi \pi \\ \nonumber
\omega N &\to & N\pi \pi \pi
\end{eqnarray}
are included with an energy independent cross section of 30 mb. 

The measured total and elastic baryon-baryon (p~+~p, p~+~n) and 
meson-baryon ($\pi^+$~+~p, $\pi^-$~+~p, $K^-$~+~p, $K^+$~+~p) 
cross sections are to a good approximation
independent of the incoming isospins for energies above
the string threshold. Paramerizations of the 
experimental  p~+~p, p~+~n, $\pi^+$~+~p, $\pi^-$~+~p, $K^-$~+~p, $K^+$~+~p 
cross sections are taken from Ref.~\cite{PartProp}. The  
cross sections for the other hadron-hadron channels 
must be specified in the transport model
for collision energies above the string threshold 
($\sqrt{s}=$ 2.65 GeV for baryon-baryon, 
$\sqrt{s}=$ 2.1 GeV for meson-baryon);
in the HSD approach these high energy cross sections  are
related to the measured cross sections by
\begin{eqnarray}
\sigma_{tot}^{N\Delta}(\sqrt{s})=0.5\left(\sigma_{tot}^{pp}(\sqrt{s})
+\sigma_{tot}^{pn}(\sqrt{s})  \right) \\ \nonumber
\sigma_{tot}^{\rho N}(\sqrt{s})=\sigma_{tot}^{\rho \Delta}(\sqrt{s})
=\sigma_{tot}^{\omega \Delta}(\sqrt{s})=
\ldots \\ \nonumber 
=0.5 \left(\sigma_{tot}^{\pi^+ N}(\sqrt{s})
+\sigma_{tot}^{\pi^- N}(\sqrt{s})\right) \\ \nonumber
\sigma_{tot}^{K N}(\sqrt{s})=\sigma_{tot}^{K \Delta}(\sqrt{s})
=\ldots  =\sigma_{tot}^{K^+p}(\sqrt{s})  \\ \nonumber
\sigma_{tot}^{\bar{K} N}(\sqrt{s})=\sigma_{tot}^{\bar{K} \Delta}(\sqrt{s})
=\ldots  =\sigma_{tot}^{K^-p}(\sqrt{s}).  \\ \nonumber
\label{cross}
\end{eqnarray}
The dots in Eq.(4) stand for all other combinations in the incoming chanels.
The same procedure is applied for the elastic cross sections. The
high energy inelastic baryon-baryon (meson-baryon) cross sections 
obtained by this procedure are $\approx$ 30 (20) mb, which correspond
to the typical baryonic (mesonic) geometrical cross sections, i.e. $\sigma^{inel}
\approx \pi R^2$. In ultrarelativistic 
heavy-ion collisions with  a lot of rescattering processes this
should be a reasonable and conservative input for the calculation.

\begin{figure}[ht]
\centerline{\psfig{figure=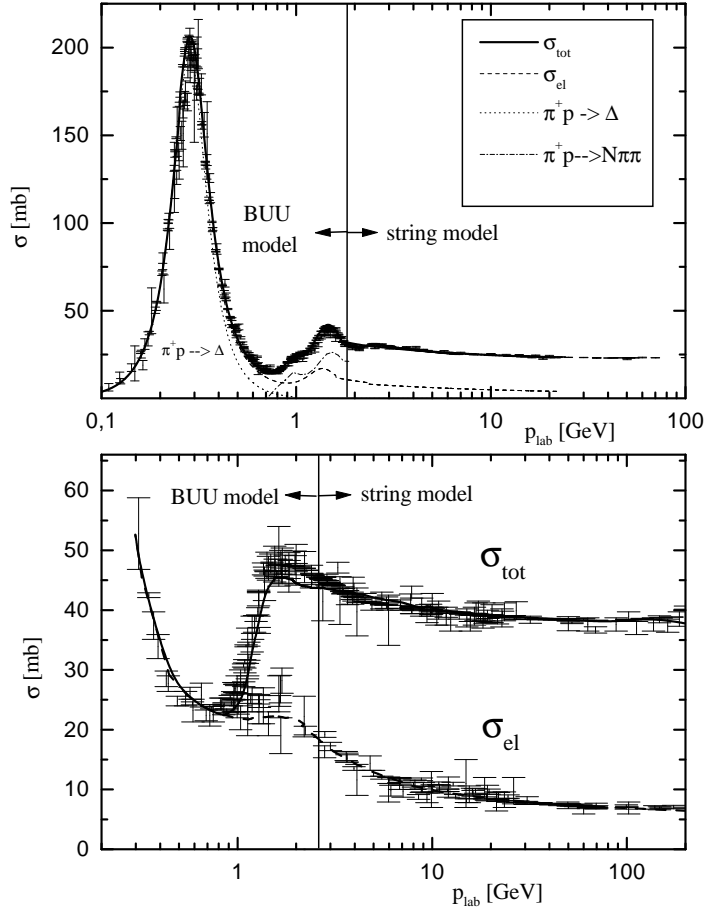,height=15cm}}
\caption{The total and elastic $\pi^+$-proton (upper part) and 
proton-proton (lower part) cross sections in comparison with the experimental
data from Ref.~\cite{PartProp}. The cross sections below the marked thresholds
are taken from the BUU model \cite{Wolf}, whereas the high energy cross
sections are parametrisations of the  experimental data.}
\label{pp-xsection}
\end{figure}

In order to be consistent
with the experimental inelastic pion-proton cross section below the string
threshold, an additional channel besides the cross section
from the BUU model \cite{Wolf} is included: $\pi N \to \pi \pi N$. 
This reaction fills up the inelastic cross section and ensures a smooth
transition at the string threshold
as shown in Fig.~\ref{pp-xsection} (upper part), where the calculated 
total and elastic $\pi^+p$ cross section together with 
the experimental data \cite{PartProp} are shown. 
The low energy nucleon-nucleon cross sections taken from the BUU model
fit reasonable well to the high energy parametrizations as shown
in Fig.~\ref{pp-xsection} (lower part) together with the
data from \cite{PartProp}. 

At ultrarelativistic (SPS) energies meson-meson reactions 
may become even more important, because the intermediate meson density is 
much higher than the baryon density. However, the average invariant
energy of these secondary or higher order reactions is rather
small. It is thus convenient to use cross sections within the 
Breit-Wigner resonance picture adopting branching ratios from 
the nuclear data tables \cite{PartProp} without introducing new parameters. 
The reactions of the type $a+b\to m_R \to c+d$, where a,b,c and d denote 
the mesons in the initial and final state and $m_R$ the mesonic intermediate
resonance ($\rho$, $a_1$, $\phi$, $K^*$), is described by the
cross section
\begin{eqnarray}
\sigma(ab\to cd)={2J_R+1 \over (2S_a+1)(2S_b+1)}\cdot {4\pi \over p_i^2}\cdot
{s \Gamma_{R\to ab}\Gamma_{R\to cd} \over (s-M_R^2)^2+s \Gamma_{tot}^2}.
\label{res}
\end{eqnarray}
In Eq.(\ref{res}) $S_a$, $S_b$ and $J_R$ are the spins of the particles; 
$\Gamma_{R\to ab}$ and $\Gamma_{R\to cd}$ denote the partial decay width
in the initial and final channels, $M_r$ and $\Gamma_{tot}$  the mass
and the total resonance width and $p_i$ is the initial momentum in the 
resonance rest frame. The following meson-meson reactions are included:
\begin{eqnarray}
\pi \pi \leftrightarrow \rho, \quad
\pi \rho \leftrightarrow \phi,\quad
\pi \rho \leftrightarrow a1,\quad
\pi K \leftrightarrow K^*.
\end{eqnarray}
Additionally strangeness production by meson-meson collision is included,
which is expected to contribute at AGS and SPS energies, where a high
mesonic density is achieved.  An isospin averaged cross section 
\cite{cassing2}
\begin{eqnarray}
\bar{\sigma}_{mm\to K\bar{K}}(s)=2.7 \cdot \left( 1-{s_0\over s}\right) ^{0.76} 
\mbox{[mb],} \quad  s_0=4 m_K^2
\end{eqnarray}
is applied, where $mm$ stand for all possible nonstrange mesons in the 
incoming channel, e.g. 
\begin{eqnarray}
\pi \pi \to K\bar{K}, \quad
\pi \rho \to K\bar{K},\quad \ldots .
\label{mmtokk}
\end{eqnarray}

\subsection{Elementary baryon-baryon collisions}
The primary elementary inelastic collisions of hadron pairs are 
the essential input for the microscopic simulation of heavy-ion collisions.
For the present analysis the elementary strangeness production in $pp$ 
collisions over a wide energy range is
of particular interest and serves as a reference for strangeness
enhancement in heavy-ion collisions. 

In the HSD approach the high energy inelastic hadron-hadron 
collisions are described by the FRITIOF model \cite{LUND}, where
two incoming hadrons will emerge the reaction as two excited
color singlet states, i.e. strings.
The energy and momentum transfer in the FRITIOF model are assumed
to happen instantanously at the collision time. With this 
phenomenological description of the soft processes the
global properties of heavy-ion collisions can be described very well
(see section \ref{Baryon stopping and pion production}).
Observables which are sensitive to hard parton-parton processes,
like e.g. Drell-Yan production or pion production with highest 
Feynman x, cannot be described within this approach \cite{geiss} (for a 
detailed discussion
of this topic see \cite{Kahana}). However, the observables
discussed in the present paper in the energy range 2-200 
A$\cdot$GeV, i.e. low to moderate transverse momentum particle production, are
dominated by the soft processes, which are described by string excitation
and fragmentation.

According to the Lund string model \cite{Anderson} 
a string is characterized by the 
leading constituent quarks of the incoming hadron 
 and a tube of color flux is supposed to
be formed connecting the rapidly receding string-ends. In the 
HSD approach baryonic ($qq-q$) and mesonic ($q-\bar{q}$) strings
are considered. 
In the uniform color field of the strings 
virtual $q\bar{q}$ or $qq\bar{q}\bar{q}$ pairs are produced 
causing the tube to fission and thus create mesons or baryon-antibaryon
pairs. The production probability $P$ 
of massive $s\bar{s}$ or $qq\bar{q}\bar{q}$ pairs
is suppressed in comparison to light flavor production 
($u\bar{u}$, $d\bar{d}$) according to a Schwinger-like formula
\cite{Schwinger}
\begin{eqnarray}
{P(s\bar{s}) \over P(u\bar{u})}=\gamma_s= 
exp\left(-\pi {m_s^2-m_q^2\over 2\kappa},
\right)
\label{schwinger}
\end{eqnarray}
with $\kappa\approx 1GeV/fm$ denoting the string tension. Thus in the 
Lund string picture the production of strangeness and 
baryon-antibaryon pairs 
is controlled by the constituent quark and 
diquark masses. Inserting the constituent quark masses
$m_u=0.3$ GeV  and $m_s=0.5$ GeV a value of $\gamma_s \approx 0.3$ is obtained.
While the strangeness production in proton-proton collisions
at SPS energies is reasonably well reproduced with this value, the strangeness 
yield for p~+~Be collisions at AGS energies is underestimated 
by roughly 30\%, as we will show
in the next section. For that reason the   
suppression factors used in the HSD model are
\begin{eqnarray}
u:d:s:uu = \left\{
\begin{array}{ll}
1:1:0.3:0.07 &, \mbox{at SPS energies} \\
1:1:0.4:0.07 &, \mbox{at AGS energies} ,
\label{ssup}
\end{array}
\right.
\label{HSD-supp}
\end{eqnarray}
with a linear transition of the strangeness suppression factor 
as a function of $\sqrt{s}$ in between. 

The production probability
for $qq\bar{q}\bar{q}$ pairs in the HSD model 
is reduced to 
\begin{eqnarray}
{P(qq\bar{q}\bar{q}) \over P(u\bar{u})}=0.07
\label{schwinger2}
\end{eqnarray}
compared to the standart FRITIOF parameter 0.1 in order to
get better agreement with $\bar{p}$ and $\bar{\Lambda}$ production
in p~+~p collisions at SPS energies as we will show in the next section. 
This parameter has no influence on the global strangeness production
in heavy-ion collisions.  

Additionally a fragmentation function $f(x,m_t)$ has to be 
specified, which is the probability distribution for
hadrons with transverse mass $m_t$ to acquire the energy-momentum
fraction $x$ from the fragmenting string
\begin{eqnarray}
f(x,m_t)\approx {1 \over x} (1-x)^a exp\left(-bm_t^2/x  \right),
\end{eqnarray}
where $a=0.23$, $b=0.34 \, GeV^{-2} $ are used in the HSD model.

In order to get some confidence about the particle production
from the FRITIOF model, we show in Fig.~\ref{pp-ags} the 
invariant cross sections for inclusive proton (upper part) and $\pi^+$ 
production in proton-proton collisions at $p_{lab}=12$ GeV  
in comparison to the data from Ref. \cite{blobel}.
The $\pi^+$ spectra are shown in the lower part 
for three different intervals of 
transverse momentum and have a Gaussian shape while the 
proton rapidity distribution is peaked around target and projectile
rapidity (upper part). The comparison with the 
data indicates a slightly too high
stopping obtained within the FRITIOF model at midrapidity, 
which one should keep 
in mind when analyzing stopping in heavy-ion collisions.
In Fig.~\ref{pp-sps} the rapidity distributions of protons (dashed line) and
negatively charged hadrons (solid line) for proton-proton
collisions at SPS energies ($p_{lab}$ = 200 GeV)  are shown. 
Again a good agreement with the data
is found in rapidity space for the produced particles. 

The particle distribution in transverse momentum is 
illustrated  in Fig.~\ref{pp-pt}a, where we show the 
$p_t$ spectra of $\pi^+$, $K^+$, $K^-$, $p$ and $\bar{p}$
for inelastic proton-proton collisions
at $\sqrt{s}=23$ GeV (SPS energies) at midrapidity $|y|\le 0.1$ in comparison
to the data from Ref.~\cite{Alber}. The calculated spectra and the data 
at midrapidity are not exponential; the pions show a clear 
low-$p_t$ and high-$p_t$ enhancement  opposite
to $p$ and $\bar{p}$ spectra. The situation is different at AGS energies
as demonstrated in Fig.~\ref{pp-pt}b, where the transverse momentum spectra of 
$\pi^+$, $\pi^-$ and $K^0_s$ for inelastic 
proton-proton collisions at $p_{lab}=12$ GeV are shown in comparison
with the data from Ref.~\cite{blobel}. 
The overall agreement of particle production, both in transverse momentum 
and in rapidity space, obtained within the FRITIOF model 
from AGS to SPS energies is rather good. 

The complete energy regime   
can be tested by data for energy dependent 
cross sections for the reactions 
$p+p \to 2\, prongs + X$, $p+p \to 4\, prongs + X$ and $p+p \to 6\, prongs + X$ 
({\em prongs}:  charged particles)
as shown in Fig.~\ref{prongs} together with the experimental data from 
\cite{Landolt}. We note that the particle production from the 
FRITIOF model, which
is taken in the HSD approach for inelastic baryon-baryon collisions, fits 
reasonably well the data from SPS energies 
down to the string threshold $\sqrt{s}=2.65$ GeV.
 
The implementation of this string fragmentation model into a covariant
transport theory implies to use a time scale 
for the particle production process, i. e.
the formation time $t_f$. The formation time includes the formation
of the string, the fission of the string due to $q\bar{q}$ and 
$qq\bar{q}\bar{q}$ production
into small substrings and the time to form physical hadrons. It can be
interpreted as the time needed for a hadron to tunnel out of the vacuum
and to form its internal wavefunction.
In the HSD model the formation time is a single fixed parameter for all
hadrons and is set to $t_f=0.8$ $fm/c$ \cite{cassing} in the rest frame of the 
new produced particle. In the center of mass of a string the hadronization 
starts after the formation time and proceeds to the stringends 
as illustrated in Fig. \ref{string_dyn}. The formation point of a new 
produced hadron with velocity $\vec{\beta}$ in the string cms is given by
\begin{eqnarray}
\vec{x}=\vec{x}_{coll}+\vec{\beta}\cdot t_f, 
\end{eqnarray}
where $\vec{x}_{coll}$ is the collision point of the two incoming 
hadrons.

Due to time dilatation and Lorentz $\gamma$-factors of $\approx 2-6$ 
for the leading 
constituent quarks for AGS to SPS energies the formation time of
the leading hadrons are long in comparison to the time between two
consecutive collisions in heavy-ion reactions. 
Thus applying the concept of string fragmentation 
to heavy-ion collisions one has to specify the interaction of strings
and their constituents with the surrounding hadrons. Here a similar 
picture as in the UrQMD model \cite{Bass} is used: 
The cross section of the secondary interactions of the leading 
quarks/diquarks are reduced prior to the formation as
\begin{eqnarray}
\sigma(q-B)&=&1/3 \,\sigma(B-B) \approx 10mb \\ \nonumber
\sigma(qq-B)&=&2/3\,\sigma(B-B) \approx 20mb \\ \nonumber
\sigma(qq-q)&=&2/9\,\sigma(B-B) \approx 6.6mb \nonumber
\end{eqnarray}
and so on. In order to treat this scheme within the FRITIOF string picture 
the $q$ ($qq$) is assumed to form a meson (baryon) together with its
prospective quark partner inside the string. This procedure
has to be seen as a heuristic approximation of the underlying soft 
partonic dynamics. Nevertheless, the global properties of heavy-ion 
collisions, the baryon stopping and pion production, can be 
described with this procedure over a wide energy range as we will
show in Section \ref{Baryon stopping and pion production}.   
 
The interaction of the string field spanned between the
constituent quarks with other hadrons is not taken into account.
This is motivated by the fact, that most of the strings in a given
space-time volume fragment within a small time intervall. Thus the
interaction of secondaries with the string field should be negligible
in first order. Furthermore, since most of the strings are stretched 
longitudinally, no string-string
interaction or a string fusion to color ropes as suggested
in \cite{Hsorge,Werner2} is included in order to avoid new parameters.

\subsection{Strangeness production in elementary hadron-hadron collisions}
\label{Strangeness production in elementary hadron-hadron collisions}
Of particular interest for the present analysis is the strangeness 
production in the elementary hadron-hadron collisions. 
To obtain agreement with p-p data at SPS energies and with
p-Be at AGS energies the strangeness suppression factor $\gamma_s$
in the FRITIOF model had to be enhanced from $\gamma_s=0.3$ (SPS) to 
$\gamma_s=0.4$ (AGS) as mentioned above. To illustrate this we show in Table
\ref{tab-sps} the particle multiplicities obtained in 
p-p collisions at $p_{lab}=200$ GeV  in comparison to the experimental 
data from \cite{Kafka}. The agreement in the strangeness sector is very good,
however, the $\Lambda+ \Sigma^0$ yield is overestimated by roughly 30\%,
but is fixed in the calculation by the $K^+$, $K^-$ and $K^0_s$ yield
due to strangeness conservation. At AGS energies the suppression factor $\gamma_s$=0.4
is choosen in order to describe the kaon production in p~+~Be collisions;
otherwise the strangeness production here would be underestimated by 30\%
(see Fig.~\ref{pbepau-strange}).

\begin{table}
\centerline{\begin{tabular}{|c|c|c|c|}
\hline
particle & data & HSD  \\
\hline\hline
$\pi^+$   & 3.22 $\pm$ 0.12 & 3.25  \\ \hline
$\pi^-$   & 2.62 $\pm$ 0.06 & 2.53  \\ \hline
$\pi^0$   & 3.34 $\pm$ 0.24 & 3.36  \\ \hline
$K^+$   & 0.28 $\pm$ 0.06 & 0.274  \\ \hline
$K^-$   & 0.18 $\pm$ 0.05 & 0.18  \\ \hline
$K^0_s$   & 0.17 $\pm$ 0.01 & 0.174  \\ \hline
$\Lambda +\Sigma^0$   & 0.1 $\pm$ 0.015 & 0.15  \\ \hline
$\bar{\Lambda}+\bar{\Sigma^0}$   & 0.013 $\pm$ 0.01 & 0.018  \\ \hline
$p$   & 1.34 $\pm$ 0.15 & 1.32  \\ \hline
$\bar{p}$   & 0.05 $\pm$ 0.02 & 0.057  \\ \hline
\end{tabular}}
\caption{Particle multiplicities for inelastic proton-proton collisions at
$p_{lab}$ = 200 GeV compared to the data from \cite{Kafka}.}
\label{tab-sps}
\end{table}

Since the string model is designed to describe inelastic hadron-hadron
collisions at rather high energies, its results for the strangeness 
production down to threshold become questionable. On the other hand 
the low energy cross sections are of particular interest for
strangeness production during the rescattering phase. For that reason
explicit parametrizations of the channel
$N N\to N Y K$  have to be used. The isospin 
averaged cross sections of  this channel are related to the measured 
channel as:
\begin{eqnarray}
\label{nnk}
\sigma_{N N\to N \Lambda K} &=& 3/2 \, \sigma_{pp\to p \Lambda K^+} \\ \nonumber
\sigma_{N N\to N \Sigma K} &=& 3/2 \, \left(
        \sigma_{pp\to p \Sigma^+ K^0}+\sigma_{pp\to p \Sigma^0 K^+}\right) .
\end{eqnarray}  
The explicit cross sections are approximated by a fit to experimental
data and are specified in Ref.~\cite{cassing2}. The same procedure is
applied to other combinations of incoming particles (p, n, $\Delta$),
where isospin averaged cross section for the low energy parametrisations
are taken (for details see  Ref.~\cite{cassing2}).

The total kaon production in p~+~p collisions from threshold up to
$\sqrt{s}=100$ GeV 
within the HSD approach is presented 
in Fig.~\ref{pp-to-k}a as a function of the
invariant energy above threshold $\sqrt{s}-\sqrt{s_0}$ together with 
the experimental data \cite{Landolt,Giao,Cosy,TOF}. In the HSD approach 
the $K^-$ are produced in baryon-baryon collisions only via string 
fragmentation, because the threshold 
$\sqrt{s_0}=2\cdot m_p+2\cdot m_k$ is above the string threshold.
On the other hand the threshold for $K^+$ production 
$\sqrt{s_0}=m_p+m_{\Lambda}+m_K=2.55$ GeV is below the string threshold
as shown in Fig.~\ref{pp-to-k}a. The low energy parametrization
of the kaon production (c.f. eq. (\ref{nnk})), 
which is also shown in Fig.~\ref{pp-to-k}a, gives a smooth
transition at the string threshold. 
Thus the kaon production in nucleon-nucleon collisions can be reproduced
within the HSD approach over many orders of magnitude.

A further important strangeness production channel are meson-baryon collisions,
which are especially important in heavy systems where the secondary
mesons are produced after the formation time $\gamma \times t_f$
inside the nuclei. The same procedure as for baryon-baryon
collisions is applied below the string threshold (2.1 GeV).
For the channels
\begin{eqnarray}
\pi N &\to &Y K \\ \nonumber
\pi \Delta &\to& Y K, 
\end{eqnarray}  
where $Y=\Lambda,\Sigma$ and $K=K^+,K^0$, we adopt the detailed parametrizations 
from Tsushima et al. \cite{tsu}.
The reaction $\pi N\to N K \bar{K}$ is also included with a 
cross section taken from Ref.~\cite{cassing2}
\begin{eqnarray}
\sigma(\pi^-p\to p K^0 K^-)=1.121 \left(1-{s_0 \over s}  \right)^{1.86}
\left({s_0 \over s}  \right)^{2} [mb]
\end{eqnarray}  
where $\sqrt{s_0}=m_N+2m_K$, which is a parametrization
of the experimental data. Using isospin symmetries \cite{cassing2}
the different isospin channels are related to $\sigma(\pi^-p\to p K^0 K^-)$.

For $\sqrt{s}>2.1$ GeV strangeness is only produced by string 
fragmentation. However, the strangeness production in meson-baryon collisions
calculated with the FRITIOF model underestimates  the 
experimental data in this energy regime by $\approx 1mb$ 
as shown in Fig.~\ref{pp-to-k}b,
where the dashed line is the FRITIOF result for the inclusive cross section
$\pi^- p \to strangeness + X$. This difference can easily be understood within
the FRITIOF string picture: The remnants of a hadron-hadron collision
are always two strings and the number of incoming constituent quarks
remains unchanged. Thus strangeness production in this string picture is
always connected with new particle production. For baryon-baryon collisions
this is reasonable, while for meson-baryon collisions it is no longer
valid because a reaction like $\pi N\to K \Lambda$  never can be described.
This corresponds to an annihilation of constituent quarks
$u\bar{u}\to s\bar{s}$, which is not included in the FRITIOF model.
For that reason we add the channel $\pi N\to K \Lambda$ 
explicitly with an energy independent  
cross section of 1 mb resulting in a better 
description of strangeness production in $\pi N$ collision. 
In Fig.~\ref{pp-to-k}b the HSD result for the inclusive cross section
$\pi^- p \to strangeness + X$ is shown by the solid line in comparison to 
the data \cite{Landolt}. 

\section{Baryon stopping and pion production}
\label{Baryon stopping and pion production}
Since the baryon and pion dynamics of nucleus-nucleus collisions 
at SIS energies has been
investigated in detail in Refs.~\cite{cassing,cassing2,cassing3} we focus on
AGS and SPS energies in this Section.

\subsection{AGS energies}
\label{AGS energies}
At AGS energies ($\leq$ 15 A$\cdot$GeV) the initial nucleon-nucleon
collisions occur at $\sqrt{s} \approx$ 5 GeV and the Lorentz
contraction of the nuclear density in the nucleon-nucleon cms amounts
to $\gamma_{cm} \approx$ 3.  Thus most of the mesons produced in p~+~Be
reactions hadronize (after their formation time $t_f \times  \gamma$) in the
vacuum without rescattering such that this light system may serve as a
test for the LUND string-model employed at $\sqrt{s} \approx$ 5 GeV. In
this respect we show in Fig.~\ref{pbepau-stopping} the inclusive proton and
$\pi^-$ rapidity spectra for p~+~Be and p~+~Au
at 14.6 A$\cdot$GeV in comparison
to the data from the E802 Collaboration \cite{E802p}. The approximate
symmetry of the $\pi^-$ rapidity distribution around midrapidity $y_{CM}$
for p~+~Be indicates very little rescattering of the pions. Also the proton
distribution is rather well reproduced by the calculation for $t_f$ =
0.8 fm/c, which we consider as the 'default' value for the universal 
formation time.

The effect of pion rescattering on nucleons and secondary pion production
channels in p~+~Au at 14.6 GeV
collisions can be extracted from the lower part of
Fig.~\ref{pbepau-stopping}  where the pion rapidity
distribution is no longer symmetric around $y_{CM}$, but sizeably enhanced
at target rapidity ($y_{lab} \approx $ 0). The stopping of protons 
(dashed line) is also
clearly visible in the proton rapidity distribution, 
both in the calculations as well as in the data of the E802
Collaboration \cite{E802p}.

\begin{figure}[ht]
\centerline{\psfig{figure=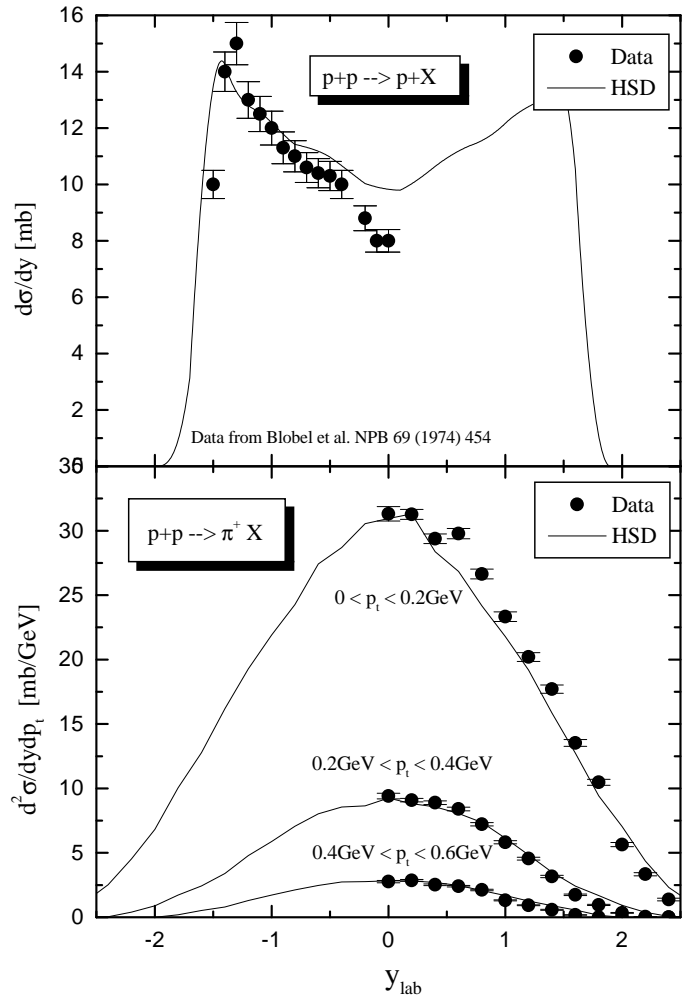,height=15cm}}
\caption{Invariant cross sections for inclusive proton (upper part) and 
$\pi^+$ production (lower part) in proton-proton
collisions at $p_{lab}=12$ GeV in comparison to the data from Ref. \cite{blobel}.
The $\pi^+$ results are shown for three different intervals 
of transverse momentum.}
\label{pp-ags}
\end{figure}

\begin{figure}[ht]
\centerline{\psfig{figure=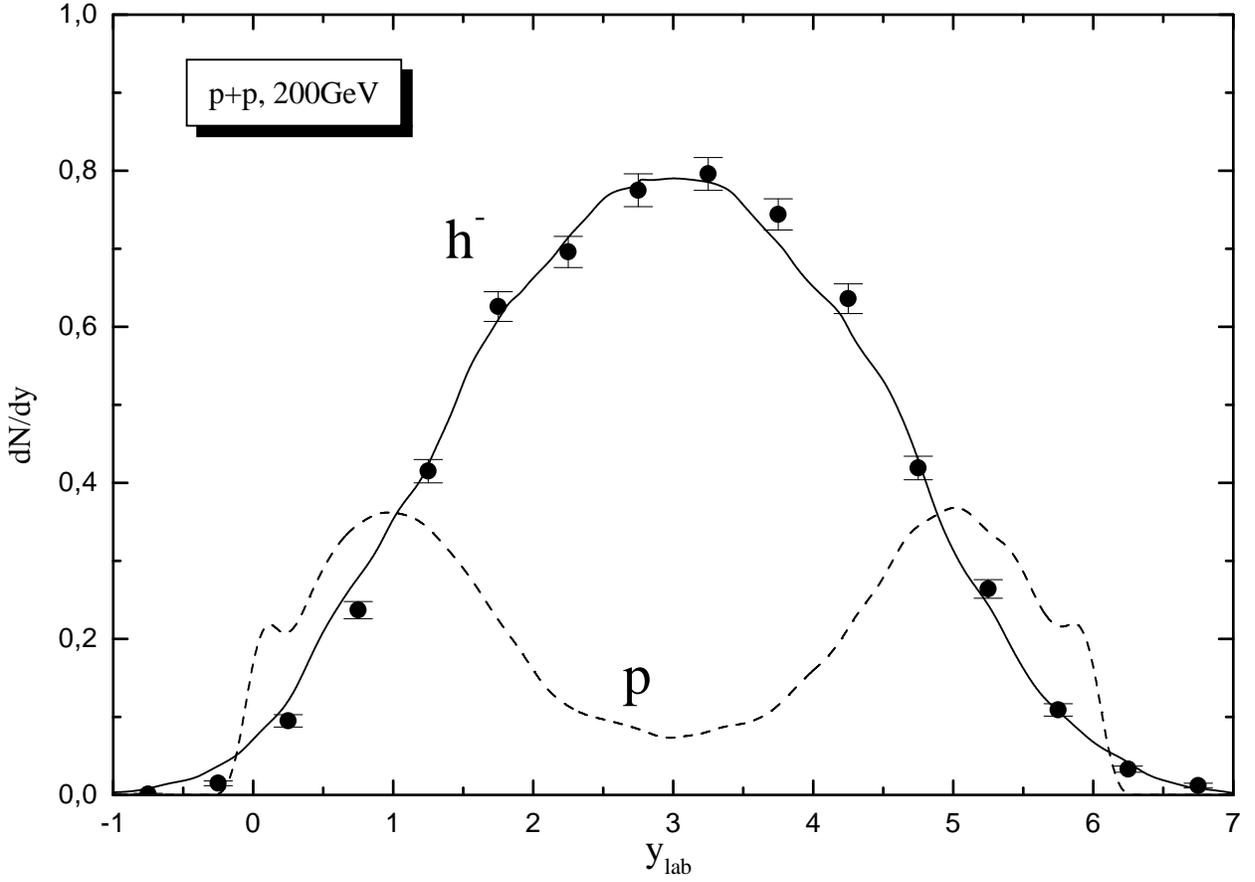,height=13cm}}
\caption{Rapidity distributions of protons (dashed line) 
and negatively charged hadrons (solid line)
in proton-proton collisions at $p_{lab}=200$ GeV in comparison 
to the data from Ref. \cite{pp-sps}.}
\label{pp-sps}
\end{figure}

\begin{figure}[ht]
\centerline{\psfig{figure=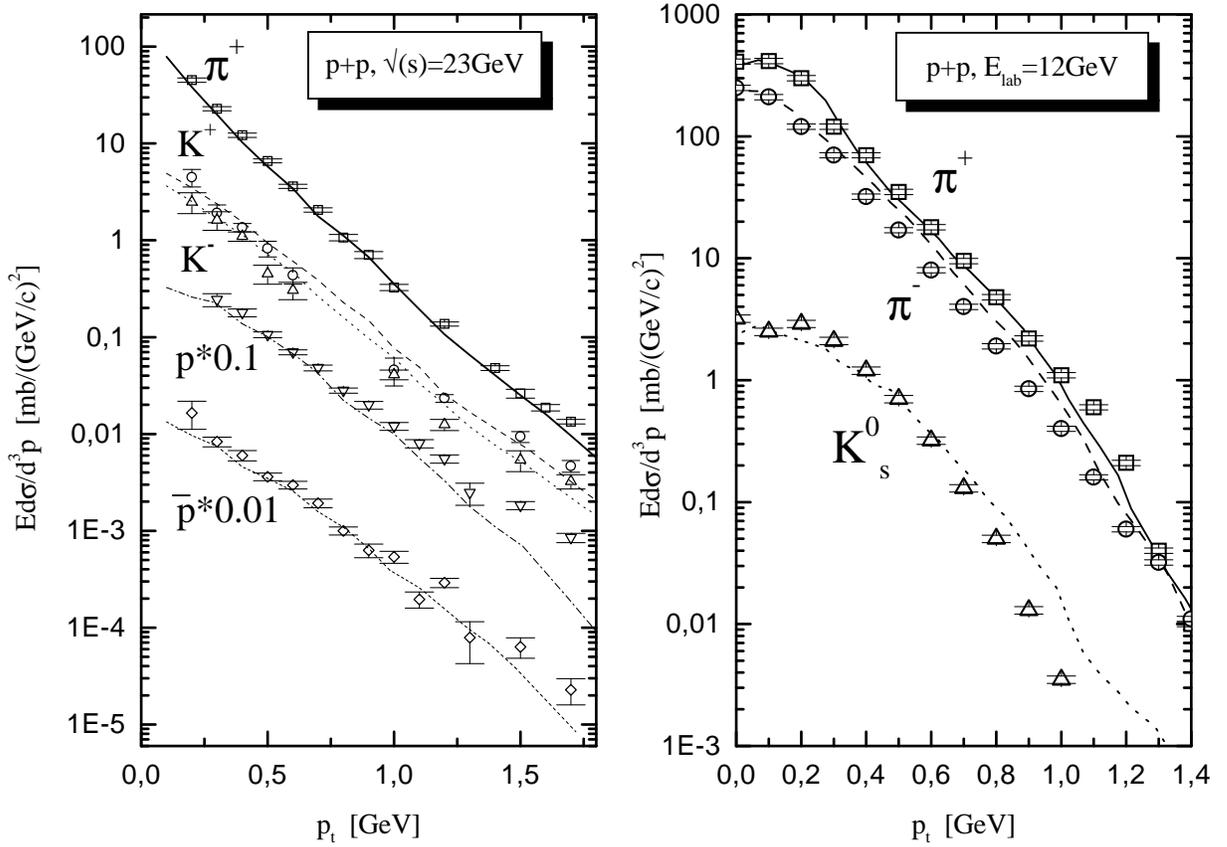,height=13cm}}
\caption{{\em l.h.s}: Transverse momentum spectra of 
$\pi^+$, $K^+$, $K^-$, $p$ and $\bar{p}$
for inelastic proton-proton collisions
at $\sqrt{s}=23$ GeV (SPS energies) at midrapidity $|y|\le 0.1$ in comparison
to the data from Ref.~\cite{Alber}. The $p$ and $\bar{p}$ spectra 
are scaled down
by factors of 0.1 and 0.01, respectively.
{\em r.h.s}: Transverse momentum spectra of 
$\pi^+$, $\pi^-$ and $K^0_s$ for inelastic 
proton-proton collisions at $p_{lab}=12$ GeV (AGS energies). The data are from
Ref.~\cite{blobel}.}
\label{pp-pt}
\end{figure}

\begin{figure}[ht]
\centerline{\psfig{figure=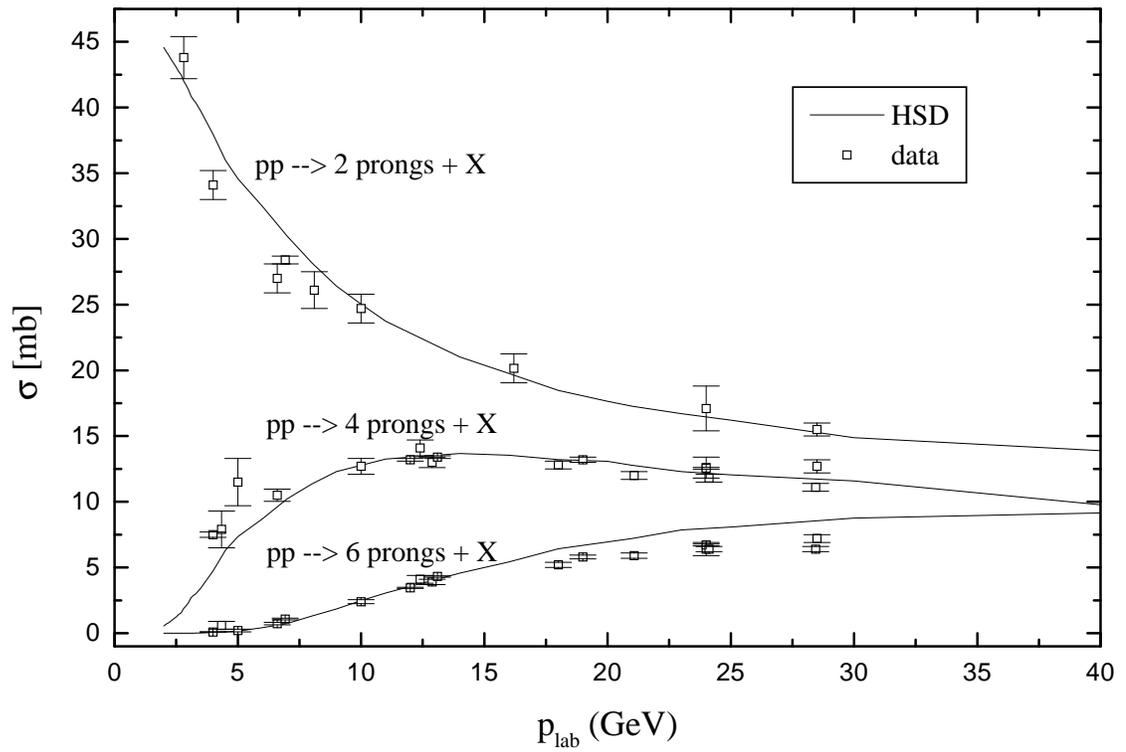,height=12cm}}
\caption{The energy dependent cross section for the reactions
$pp\to 2 prongs + X$, $pp\to 4 prongs + X$ and $pp\to 6 prongs + X$ in
comparison to the experimental data from \cite{Landolt}.} 
\label{prongs}
\end{figure}

\begin{figure}[ht]
\centerline{\psfig{figure=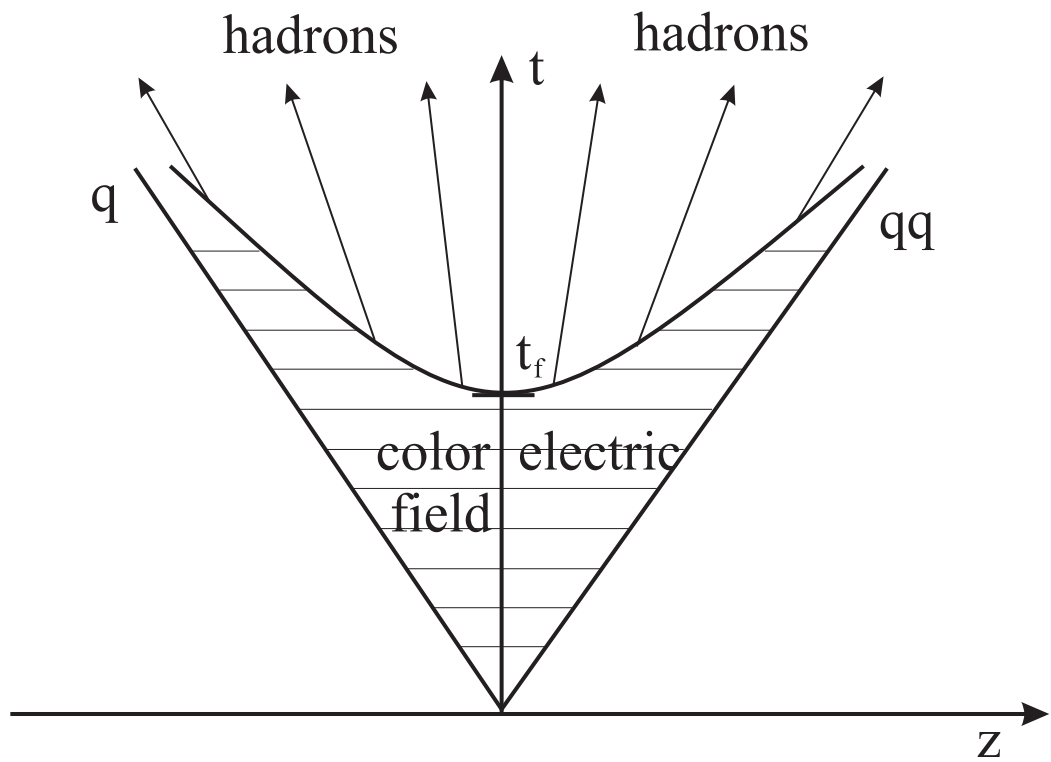,height=10cm}}
\caption{Dynamical evolution of a baryonic string; the fragmentation
into hadrons starts after the formation time $t_f$.}
\label{string_dyn}
\end{figure}

\begin{figure}[ht]
\centerline{\psfig{figure=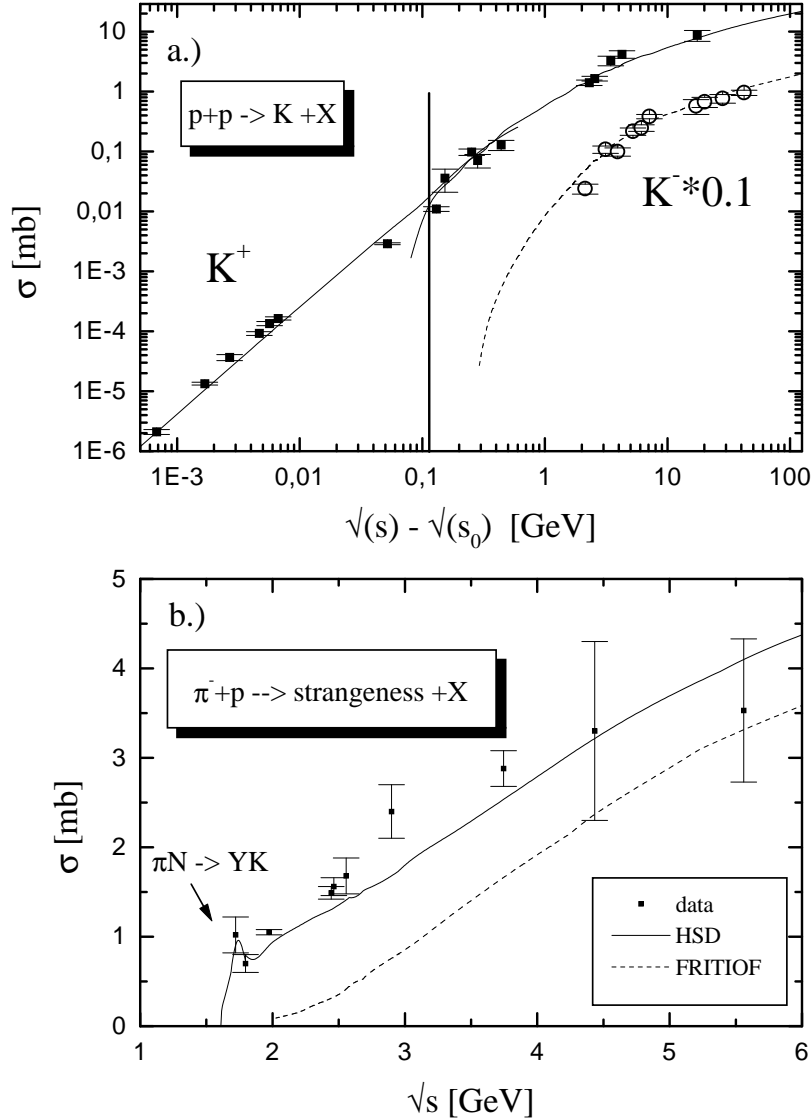,height=17cm}}
\caption{a.) Inclusive cross section for $K^+$ (full line) and $K^-$ 
(dashed line, 
$\times$0.1)
production in p~+~p reactions as a function of the
invariant energy above threshold in
comparison to the data \cite{Giao,Cosy,TOF,Landolt}. 
For $K^+$ production the string threshold is marked by the solid line; 
below the threshold ($\approx$0.11 GeV)
the parametrisations from Ref.~\cite{cassing2} are included in the HSD code.
b.) Inclusive strangeness production in $\pi^-$~+~p collisions 
in comparison with the data
from \cite{Landolt}.}
\label{pp-to-k}
\end{figure}

\begin{figure}[ht]
\centerline{\psfig{figure=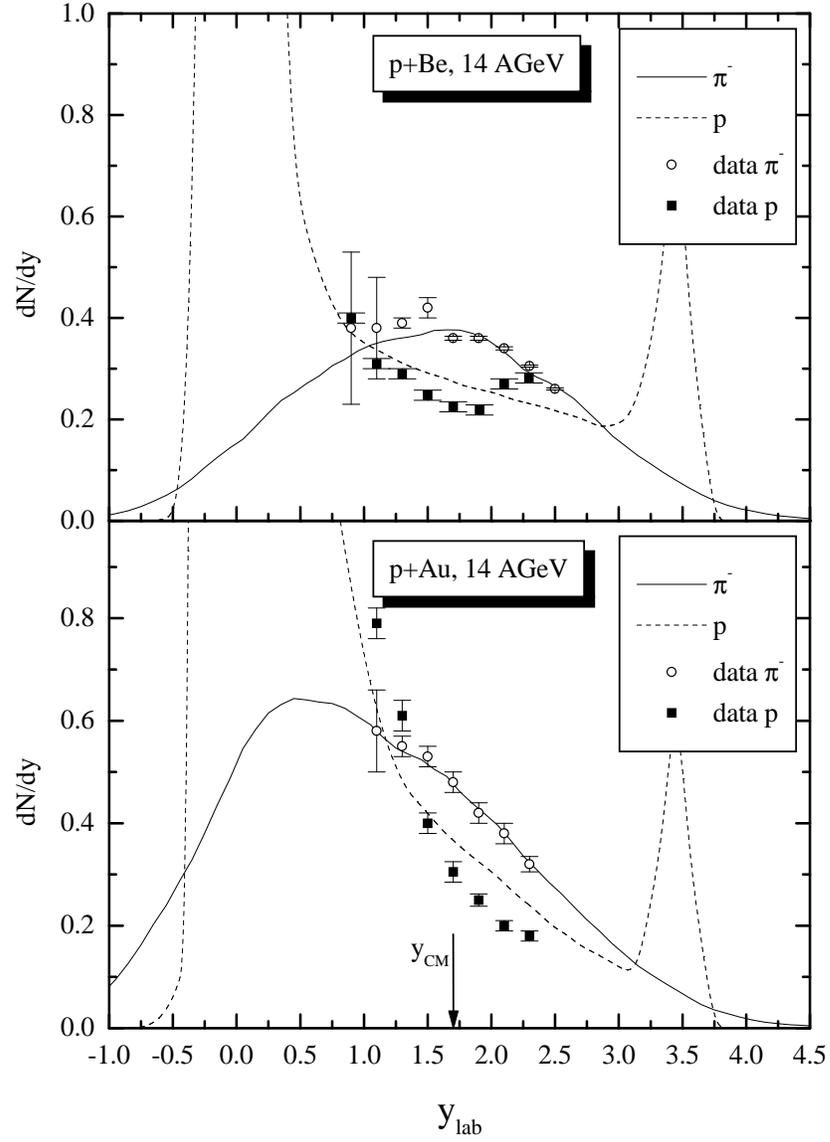,height=18cm}}
\caption{Inclusive proton and $\pi^-$ rapidity spectra for p~+~Be (upper part) 
and p~+~Au (lower part) at 14.6 A$\cdot$GeV in comparison 
to the data from the E802 Collaboration \cite{E802p}.}
\label{pbepau-stopping}
\end{figure}

The next system addressed is Si~+~Al at 14.6 A$\cdot$GeV.  The computed
rapidity distribution of protons and $\pi^-$-mesons for central collisions 
(b $\le$ 1.5 fm) is
compared in Fig.~\ref{AB-ags-stopping}a to the data from Ref.~\cite{abbott}.
Whereas the proton rapidity distribution turns out to be quite flat in
rapidity $y$ due to proton rescattering, the pion rapidity distribution
is essentially of Gaussian shape which reflects the pion rapidity
spectrum from the string fragmentation model 
(cf. Fig.~\ref{pp-ags}).  We note, however, that the width of the
pion rapidity distributions in the HSD approach is wider compared
to the E802 data. In the calculation the full width 
at half maximum (FWHM) ($\approx 2.8$ ) is only tiny 
lowered compared to  p~+~p collisions ($\approx 2.9$, c.f. 
Fig.~\ref{pp-ags}) due to the small amount of rescattering
in Si~+~Al. The maxima in the calculated rapidity spectra at target and
projectile rapidity do not show up in the data due to acceptance cuts. 

In analogy to Fig.~\ref{pp-pt} we show in Fig.~\ref{sial-pt} the
calculated transverse mass-spectra of $\pi^-$-mesons for Si~+~Al at
14.6 A$\cdot$GeV (solid lines) in comparison to the experimental data
from Ref.~\cite{abbott}. The overall agreement for  rapidities of
$y_{lab}=$ 0.5, 0.7, 0.9, 1.1, 1.3 seems to indicate that the general reaction
dynamics for pions is rather well reproduced within the HSD approach,
although the $\pi^-$ rapidity spectrum is slightly narrower in
experiment as compared to the calculations.

Nucleon stopping becomes more pronounced for the system Si~+~Au at 14.6
A$\cdot$GeV as seen from Fig.~\ref{AB-ags-stopping}b where the calculated
proton and $\pi^-$ rapidity distributions (for $b \leq$ 3.5 fm) are
compared to the data from E802 \cite{abbott} (full squares).  
Whereas the proton stopping is
reasonably well reproduced by the calculation the pion spectra again clearly
come out too broad in rapidity  as compared to the experimental data.
The overestimate of pions in the HSD approach in Si~+~Au collision
is in agreement with the findings within the RQMD model \cite{Gonin} and 
the ARC model \cite{ARC}. Thus the experimentally observed strong 
reduction in the width of the pion rapidity
spectra in heavy-ion collisions compared to p~+~p collisions at AGS 
energies, already observed in the light system Si~+~Al,
seems to be a general problem for microscopic hadronic models.

The amount of stopping at AGS energies is most clearly pronounced for
central Au~+~Au reactions as displayed in Fig.~\ref{AB-ags-stopping}c for the
proton and $\pi^-$ rapidity distributions in comparison to the
experimental data from Refs.~\cite{E866,E877}. Though the pion rapidity
spectrum - which again comes out slightly too broad in the calculation -
does not differ very much in shape from that of the Si~+~Al
system in Fig.~\ref{AB-ags-stopping}a at first sight, 
the baryon distribution in momentum space 
for Si + Al is far from kinetic equilibrium whereas that for Au+Au 
at 11 A$\cdot$GeV  shows a clear approach versus equilibration 
(cf. \cite{cassing}). 
We note that the proton rapidity spectrum for central
Au~+~Au collisions at this energy shows a similar amount of stopping as
the RQMD approach \cite{Hsorge}, the ART calculations by Li and Ko
\cite{LiPRC} or the ARC calculations by Kahana et al. \cite{ARC}.

\subsection{SPS energies}
\label{Protons and pions at SPS energies}
We continue our comparison to experimental data with the system S~+~S
at 200 A$\cdot$GeV, i.e. the SPS regime. In  Fig.~\ref{AB-sps-stopping} 
(l.h.s.) we
show the proton and negatively charged  hadron (essentially $\pi^-$) rapidity
distributions in comparison to the experimental data from
\cite{NA35}. Though the experimental proton and $h^-$ rapidity
spectra are approximately reproduced, we cannot conclude on the general
applicability of our approach at SPS energies, because also more simple
models like HIJING or VENUS -- with a less amount of rescattering --
can reproduce the data in a similar way \cite{Gyulassy,Werner}. This is due to
the fact that at 200 A$\cdot$GeV the Lorentz contraction in the cms
amounts to $\gamma_{cm} \approx$ 10 such that hadronization essentially
occurs in the vacuum again and rather {\em little} rescattering occurs in S~+~S
collisions. We note that the width of the pion rapidity spectrum in S~+~S
($\approx 3.8$) is comparable to p~+~p collisions (c.f. Fig.~\ref{pp-sps})
in contrast to the experimental findings at AGS energies as mentioned in
the previous Section.

The transverse momentum spectra of negatively charged hadrons  for central 
S~+~S reactions at 200 A$\cdot$GeV are shown in Fig.~\ref{ss-pt-hm}
in comparison to the data in the cm rapidity interval
$0.8 \le y \le 2.0$ and $2.0 \le y \le 3.0$ from \cite{NA35}. 
The agreement between the data and the
HSD calculations is sufficiently good such that the baryon and pion
dynamics for the system S~+~S  is reasonably well under control.

The next system of our considerations are  
central S~+~Au reactions at 200 A$\cdot$GeV.
In Fig.~\ref{AB-sps-stopping} (middle) 
the $\pi^-$ and proton rapidity distributions in comparison to the data
from \cite{Bauer,Santo} are shown. Here the proton rapidity spectrum shows a
narrow peak at target rapidity ($y_{CM} \approx$ -3.03) which is easily
attributed to the spectators from the Au target. The bump at $y_{CM}
\approx$ -2 is mainly due to rescattering of target nucleons.  Note
that there is no longer any yield at projectile rapidity ($y \approx$
3.03) which implies that all nucleons from the projectile have
undergone inelastic scatterings.  Furthermore, around midrapidity the
$\pi^-$ distribution is large compared to the proton distribution.

Baryon stopping is most clearly seen for the system Pb~+~Pb at 160
A$\cdot$GeV.  In Fig.~\ref{AB-sps-stopping} (r.h.s) we show the proton and $h^-$
rapidity distributions in comparison to the data from NA49 \cite{NA49}.
Our computed proton rapidity spectrum for central collisions ($b \leq $
2.5 fm) is rather flat at midrapidity. It shows no dip as 
the HIJING \cite{Gyulassy} or VENUS \cite{Werner} simulations , and is not peaked
at midrapidity as compared to RQMD simulations \cite{Hsorge}. Thus full
stopping is not achieved at SPS energies even for this heavy system.
On the other hand, the $h^-$ rapidity distributions are very similar to
the S~+~S case, however, enhanced by about a factor of 6.5 $\approx$ 208/32.

Summarizing this Section,  the proton and pion rapidity distributions
and transverse pion spectra look reasonably well for the
systems studied experimentally at AGS and SPS energies. Nevertheless,
we note that the HSD pion rapidity spectra at AGS energies are slightly too
broad compared to data. The width of the pion spectra for heavy-ion collisions
calculated within the HSD model is only slightly decreased compared to
p~+~p collisions due to rescattering. The calculated FWHM of rapidity spectra 
at AGS energies changes form 2.9 (p~+~p) over 
2.8 (Si~+~Al) to 2.6 (Au~+~Au) and for SPS energies from
3.9 (p~+~p) over 3.8 (S~+~S) to 3.8 (Pb~+~Pb) as expected 
within an independent string scenario, 
while experimentally a stronger decrease is observed at AGS energies.  

\section{Strangeness production}
\subsection{AGS energies}
\label{strangeness-ags}
Previous investigations of strangeness production up to 2 A$\cdot$GeV
within the HSD model \cite{cassing2,cassing3}
have given evidence especially for antikaon potentials in the
medium due to the strong increase of the elementary production cross
section with the excess energy $\sqrt{s}-\sqrt{s}_{thres}$ 
\cite{cassing2}. While the antikaon yield was enhanced considerably 
by the potentials, the kaon abundancies were found to be affected only slightly
due to a small repulsive kaon potential.

Here we extend the investigations about strangeness production to 
AGS energies (10 - 15 A$\cdot$GeV), where the invariant energy in first chance $NN$
collisions is  $\sqrt{s} \approx$ 5 GeV, which is far above threshold
and where the production cross section changes only smoothly with energy
(cf. Fig. 7).
Furthermore, due to higher meson densities also meson-meson reaction
channels will become important especially for heavy systems such as
Au~+~Au.

We will investigate the same systems as in Section 3.1 where we have
concentrated on proton and pion rapidity distributions that we were
found to be reasonably in line with the HSD transport calculations. We start
with p~+~Be at 14.6 GeV and display in Fig.~\ref{pbepau-strange} (l.h.s.) the
calculated $K^+$ and $K^-$ rapidity distributions in
comparison to the data of the E802 Collaboration \cite{E802p}. Both
$K^+$ (upper line) and $K^-$ rapidity distributions (lower line)
are almost symmetric around
midrapidity indicating little reabsorption of both mesons due to the
small size of the target.  The $K^+$ and $K^-$ spectra are described
quite well by the
calculation using $\gamma_s$ = 0.4 (solid lines). 
We also show in Fig.~\ref{pbepau-strange} (l.h.s.) the results
calculated with a strangeness suppression factor $\gamma_s$=0.3 
(dotted lines) taken for 
the hadronization of the strings, which underestimates the data by roughly 30\%. 
For that reason $\gamma_s$=0.4 is taken in the HSD approach as mentioned 
in Section 
\ref{Strangeness production in elementary hadron-hadron collisions}
in order to explain essentially p~+~p (or p~+~Be) reactions as input.

The calculated $K^+$ and $K^-$ spectra for p~+~Au at 14.6 A$\cdot$GeV -
shown in Fig.~\ref{pbepau-strange} (r.h.s.) - are no longer symmetric around
midrapidity due to rescattering and especially $K^-$ absorption on
target nucleons.  Both spectra from
the E802 Collaboration \cite{E802p} are not really well described by the
transport approach. The $K^+$ yield, slightly enhanced compared
to p~+~Be, is underestimated by $\approx 5-10$\% whereas the $K^-$ yield is slightly
overestimated for $y_{lab} \geq$ 1.6.  
This already might indicate that kaon production at AGS energies
is not perfectly understood for p~+~A reactions. 

How does the situation look like in light nucleus-nucleus collisions?
The $K^+$ and $K^-$ rapidity distributions for central Si~+~Al reactions at 14.6
A$\cdot$GeV are shown in Fig.~\ref{ags-strange} (l.h.s.) in comparison
to the data from \cite{abbott}. Here the $K^+$ yield as well as
the $K^-$ yield are {\em underestimated} by roughly 20-30\%!
In this rather light system there is only a small amount of 
rescattering, since most secondary particles are produced
(after their formation time) in the vacuum. This indicates
that the primary production mechanism of kaons and antikaons is not sufficiently
described within our hadronic model!

The situation becomes worse for Si~+~Au at 14.6 A$\cdot$GeV as shown in
Fig.~\ref{ags-strange} (middle) where our calculations 
underpredict the $K^+$ and $K^-$ rapidity distributions
significantly from E802 and E859 \cite{abbott,Cole}.
The situation is similar for the RQMD approach for Si~+~Au as
demonstrated in Fig.~8 of Ref. \cite{Gonin}.
Whereas RQMD also overestimates the $\pi^+$ and $\pi^-$ spectra slightly
\cite{Gonin}, the $K^+/K^-$ ratio and especially the
$K^+/\pi^+$ ratio is underestimated sizeably in comparison
to the data from E802.  Since strangeness is conserved
in both calculations (RQMD and HSD) we have to conclude here that the
initial production of strangeness, i.e. $\bar{s} s$ pairs, is
underestimated in the hadronic models.

The heaviest system studied at the AGS is Au~+~Au at $\approx$ 11
A$\cdot$GeV.  Our calculated kaon and antikaon rapidity spectra for
semicentral (5-12\%) reactions (1.5 fm $\leq b \leq$ 3.0 fm) 
are displayed on the r.h.s.
in Fig.~\ref{ags-strange} in comparison to the data from Ref.
\cite{Ogilvi}. Again we underestimate the kaon yield by roughly
30 \%. We note that this is also the case
for most central collisions, which are not shown here.

In view of the systematic presentation of our results in comparison to
data from p~+~Be to Au~+~Au collisions we infer that the hadronic transport model
does not accurately enough describe the strangeness production in
these systems as to allow for definite conclusions. 
Furthermore, independent calculations within the ART-code
from Li and Ko \cite{LiPRC} seem to describe the $K^+$ spectra for
central Au~+~Au reactions.
Unfortunately, the latter calculations have not been applied to the
other systems (p~+~Be, p~+~Au, ...) presented here,  such that in case of conflicting results
between different transport calculations no unbiased message can be
extracted.

\subsection{SPS energies}

Since about 2 decades the strangeness enhancement in ultrarelativistic
nucleus-nucleus collisions has been proposed as a possible signature
for the formation of a quark-gluon-plasma (QGP) \cite{Rafelski1}. However,
strangeness is produced also in all energetic collisions of nonstrange
mesons with nonstrange baryons as well as nonstrange meson-meson
collisions. Thus the relative abundance of these secondary and ternary
reaction channels will be of delicate importance in determining the relative
$\bar{s} s$ enhancement compared to $pp$ collisions at the same energy.
In this respect we display in Fig.~\ref{bb-mb} the number of $BB$
and $mB$ collisions as a function of the invariant collision energy
$\sqrt{s}$ for central S~+~S collisions at 200 A$\cdot$GeV and Pb~+~Pb collisions
at 160 A$\cdot$GeV. The baryon distributions for both systems show
moderate peaks around the initial $\sqrt{s} = (4 m_N^2+2 m_N
T_{lab})^{1/2}$, but extend over the whole $\sqrt{s}$ regime with
an even more pronounced peak slightly above $\sqrt{s} = 2 m_N$.  Whereas the
first peak corresponds to the first-chance nucleon-nucleon collisions,
the latter one represents low energy comover scattering. 
Most of the intermediate BB collisions are secondary or higher order
collisions of the leading constituents of the strings, which are included
in the BB distributions. Due to the
larger size of the system these intermediate energy $BB$ collisions
are  enhanced for Pb~+~Pb as compared to S~+~S. 
Meson-baryon collisions and meson-meson collisions (not shown) are
about factors of 2 and 4, respectively, higher in Pb~+~Pb as compared
to S~+~S.  In view of the strangeness production threshold in $mB$
reactions of 1.612 GeV for kaons and 1.932 GeV for $K \bar{K}$ pairs,
respectively, still a considerable part of secondary $mB$ reactions can
contribute to the net strangeness production.

Since the Lund-string-model (LSM) describes the strangeness production
in $pp$ collisions very well -- as illustrated in Section
\ref{Strangeness production in elementary hadron-hadron collisions} --
and also the low energy production channels are reasonably well under
control (cf. Ref. \cite{cassing2}), it is now a quantitative question if a
hadronic model will be able to describe the strangeness production
in proton-nucleus and nucleus-nucleus collisions at SPS energies. 
In Fig~\ref{pA-SPS} we compare the HSD results to early experimental measurements
of p~+~S and p~+~Au reactions by the NA35 Collaboration \cite{NA35pA}. 
The p~+~S collisions 
are not truly minimun bias as p~+~Au, but require a minimun of
five charged particles to be detected by the NA35 streamer chamber.
The hyperon and $K^0_S$ distributions for p~+~A collisions follow
the trend in the experimental data. However, the calculated   
$K^0_S$ rapidity distribution for the p~+~Au system is broader compared
to the data and shows no peak at target rapidity. These findings are similar
to results obtained within the RQMD model \cite{Sorge3}, although 
RQMD gets slightly more kaons and hyperons for p~+~A collisions, which may be
attributed to a different treatment of the production and interaction
of resonances at intermediate energies.

Let us turn now to heavy-ion collisions. Our results for central 
collisions of S~+~S at 200 A$\cdot$GeV for the rapidity
distributions of $K^+$ and $K^-$  are displayed in Fig~\ref{sps-strange}
in comparison to the data from the NA35 \cite{bachles,ToporPop} and 
NA49 Collaborations \cite{Bormann}.
Since these data as well as the corresponding pion rapidity
distributions (cf. Fig. \ref{AB-sps-stopping}) are described quite reasonably
in the hadronic transport approach, the quoted strangeness
enhancement can also be explained in a hadronic scenario
including rescattering.  This has been pointed out by Sorge
since a couple of years \cite{Sorge}; our independent
calculations thus support his findings. 

For the light system S~+~S about 90\% of $K^+$ and 82\% of $K^-$ stem from
$BB$ collisions whereas the contribution from $mB$ reactions is 6\% for
$K^+$ and 8\% for $K^-$; 4\% of $K^+$ arise from $mm$ reactions and
about 7\% of $K^-$ mesons. The residual $K^-$ seen asymptotically stem
from $\pi Y$ channels.  Since in Pb~+~Pb collisions these secondary and
ternary reactions are more frequent, one might expect an even stronger
enhancement of strangeness production for the heavier system.

Our results for the kaon and antikaon rapidity distributions
for central collisions
of Pb~+~Pb at 160 A$\cdot$GeV are shown in Fig.~\ref{sps-strange}
(r.h.s.) in comparison to the data from \cite{Bormann}. 
As for S~+~S at 200 A$\cdot$GeV the $K^+,
K^-$ distributions are reproduced rather well.
The calculated $\Lambda$ rapidity distribution (dotted line in 
Fig.~\ref{sps-strange}) is fixed by the kaon yield due to strangeness
conservation but it underestimates slightly the NA49 data \cite{Bormann} 
(open triangles) at midrapidity. 

Contrary to S~+~S reactions the kaon production by $mB$ channels for Pb + Pb
collisions increases to about 20\% and $mm$ channels give roughly 15\% in case  of
$K^+$ mesons. Antikaons, that are detected finally, stem from $BB$
collisions by $\approx$ 52\%, further 20\% come from $mB$ reactions,
13\% from $mm$ channels and 15\% from $\pi Y$ channels which indicates
the relative importance of secondary and ternary reactions for the
heavy system. Thus the strangeness enhancement in the kaon and $\Lambda$-particle 
sector seen at SPS energies appears to be compatible with a hadronic reaction 
scenario.

\section{Stepping down to SIS energies}
A further test of strangeness production is to study
the energy range between AGS and SIS energies (1-2 A$\cdot$GeV). At SIS 
energies the HSD approach describes the data for kaon ($K^{+}$) 
production rather well
\cite{cassing2,cassing3} even without including selfenergies, since the
kaon potential should only be slightly repulsive.
To illustrate this we show 
in Fig.~\ref{nini} the $K^+$ rapidity spectra
for Ni~+~Ni collisions at 1.93 AGeV in comparison
to the data from the FOPI Collaboration \cite{FOPI}.  As in 
Ref.~\cite{cassing3} the $K^+$ yield is well described without including
any medium effects for the kaons.

The E866 and E895 Collaborations recently 
have measured Au~+~Au collisions
at 2,4,6 and 8 A$\cdot$GeV kinetic energy at the AGS. Thus it is of 
particular interest
to look for a {\em discontinuity in the  excitation functions} 
for pion and kaon rapidity distributions and to
compare them to our hadronic model. 
In Table~\ref{tab4} the $K^+/\pi^+$ ratios for central (b=2 fm) Au~+~Au
collisions at 2,4,6,8 and 11 AGeV are shown together with
the preliminary data. The ratio at midrapidity 
${|y-y_{CM}|\over y_{CM}} < 0.25$ is slightly higher, because the kaon
rapidity distribution is narrower than that of the pions.
While the scaled kaon yield at 2 AGeV
(SIS energies) is described in the HSD approach within the 
experimental errorbars, the experimental $K^+/\pi^+$ ratio at 4 AGeV 
is underestimated 
already by a factor of 2 and saturates at roughly 19\% for 11 AGeV.

Recent RQMD calculations \cite{Hsorge2,Ogilvi} obtain 
higher $K^+/\pi^+$ ratio in the energy range 2-11 A$\cdot$GeV 
for Au~+~Au collisions, which are even above the experimental values.
The strong increase compared to earlier RQMD versions \cite{Gonin} 
results from some conceptual new steps in the meson-baryon sector 
as described in \cite{Hsorge}.
As mentioned before, an analysis for different systems over
the complete energy range would be helpful for a quantitative
comparison of the different approaches.  

The difference of the calculated scaled kaon yield and the data for
the Au~+~Au collisions at 2-11 AGeV 
may be connected with an overestimate of the pion yield, an 
underestimate of the kaon yield or an admixture of both 
in the HSD approach. In this context we show in 
Fig.\ref{auau-excitation} our predictions of the excitation
function of the $\pi^+$ and $K^+$ rapidity distributions
for central (b=2 fm) Au~+~Au collisions at 2,4,6,8 and 10 A$\cdot$GeV.
A comparison with data -- to come up in the near future -- 
will clarify the question about
pion excess and/or missing strangeness.

\section{Summary}
We have presented a systematical study of strangeness production
from SIS to SPS energies for different systems using the
HSD approach in the cascade mode. An important ingredient of the 
present analysis
are the elementary cross sections for strangeness production
in baryon-baryon, meson-baryon and meson-meson channels, which have been 
discussed in detail and are in good agreement with the experimental data. 
In order to avoid (partly unknown) parameters, neither vector or scalar
fields are included for the baryons nor any selfenergies of mesons. 
In this respect our investigations have to be seen as a rather conservative
approach. Furthermore, only
baryon resonances up to $N^*(1535)$ are included 
since most of the properties of higher resonances
in dense hadronic matter are unknown, in particular their decay into
the strangeness sector ($N^*\to KY$ or $m^*\to K\bar{K}$) and their 
width, which are expected to be broadened substantially in the hadronic 
environment. In the present
approach the corresponding excitations of baryons and mesons
are described by strings. Their decay into the strangeness channels
and especially the  $K/\pi$ ratio in the final state is given by a 
single 
parameter $\gamma_s$ and fixed to experimental p~+~p and $\pi^+$~+~p data.

We found an enhancement of the scaled kaon yield in heavy-ion collisions
due to hadronic rescattering both with increasing system size and energy. 
It should be emphasized that this is expected within any hadronic model:
After the primary string fragmentation the hadronic fireball starts with
a $K^+/\pi^+$ ratio far below chemical equlibrium with
$\approx 6\%$ ($\approx 8\%$) at AGS (SPS) energies before 
the hadronic rescattering starts. The average kinetic energy
and the particle density increases monotonically 
with incoming kinetic energy of the 
projectile while the life time of
the fireball increases with the system size. Thus a
smooth and continuous enhancement is expected within a hadronic model by
these effects.

To summarize our results at AGS energies,
we show in Table \ref{tab2} the calculated $K^+$ and $\pi^+$ yield
(integrated over the full rapidity space)
as well as the $K^+ / \pi^+$ ratio from HSD for 
different systems in comparison to the experimental ratios
from Ref.~\cite{E802p}. Allready for Si~+~Al 
the HSD approach overestimates the 
pion yield by roughly 15\%, as shown in Section \ref{AGS energies},
and underestimates the kaon and antikaon yield by $\approx$ 30\%.
Thus the calculated scaled kaon yield $K^+/\pi^+$ is essentially
too low as shown in the last two columns of Table~\ref{tab2}.
While the experimental ratio increases by a factor of 3 from
p~+~p to central Au~+~Au collisions, the HSD appoach gives
only a factor of 1.5. Our conclusions are similar to the findings in the 
RQMD model \cite{Gonin} for the system Si~+~Au,
where a ratio E802/RQMD of 0.83 (in the rapidity interval $0.6<y<2.8$) 
for $\pi^+$ and 1.22 for $K^+$  ($0.6<y<2.2$) was found. 
This corresponds to a  ratio $(K^+/\pi^+)_{RQMD}\approx 11.5\%$,
which is slightly higher than the HSD result. This may be connected with
different rapidity cuts, because the pion rapidity distribution is 
broader than that of the kaons. Our value of 9 \% is obtained by integrating
over full rapidity space while the values of RQMD are taken 
around midrapidity. The difference between the
resonance picture of RQMD and the string picture of HSD seems to
be rather small; however, there is a sizeable discrepancy between
the cascade results (of RQMD and HSD) and the data. 

\begin{figure}[ht]
\hspace{-2cm}
\centerline{\psfig{figure=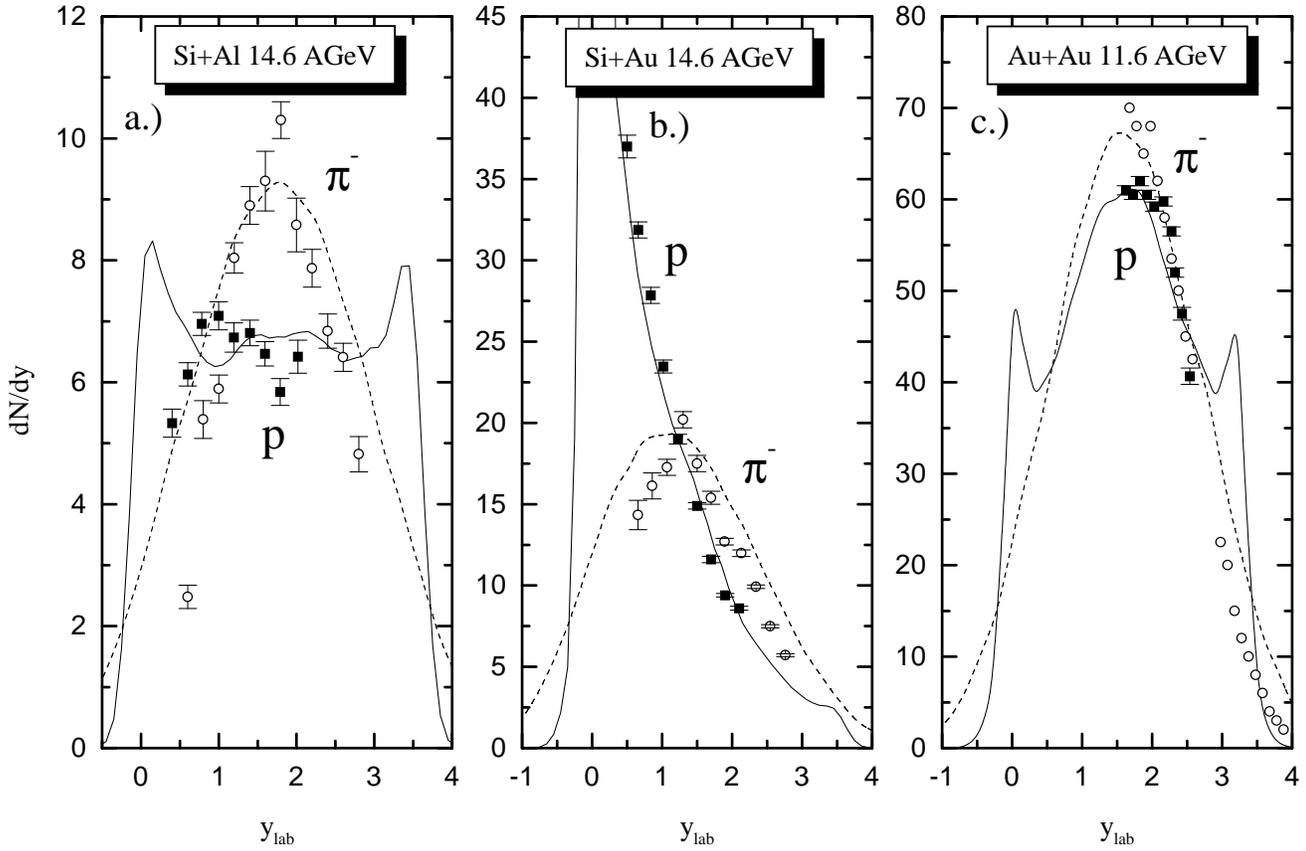,height=13cm}}
\caption{Inclusive proton (solid lines) and $\pi^-$ 
(dashed lines) rapidity spectra for central Si~+~Al ($b\le 1.5$ fm)
at 14.6 A$\cdot$GeV (left), central Si~+~Au ($b\le 3$ fm) at 14.6 A$\cdot$GeV (middle) 
and central Au~+~Au ($b\le 3$ fm) at 11.6 A$\cdot$GeV (right)
in comparison to the data from the E802 Collaboration \cite{abbott} (for Si~+~Al 
and Si~+~Au) and from the E866 and E877 Collaborations \cite{E866,E877}.}
\label{AB-ags-stopping}
\end{figure}

\begin{figure}[ht]
\centerline{\psfig{figure=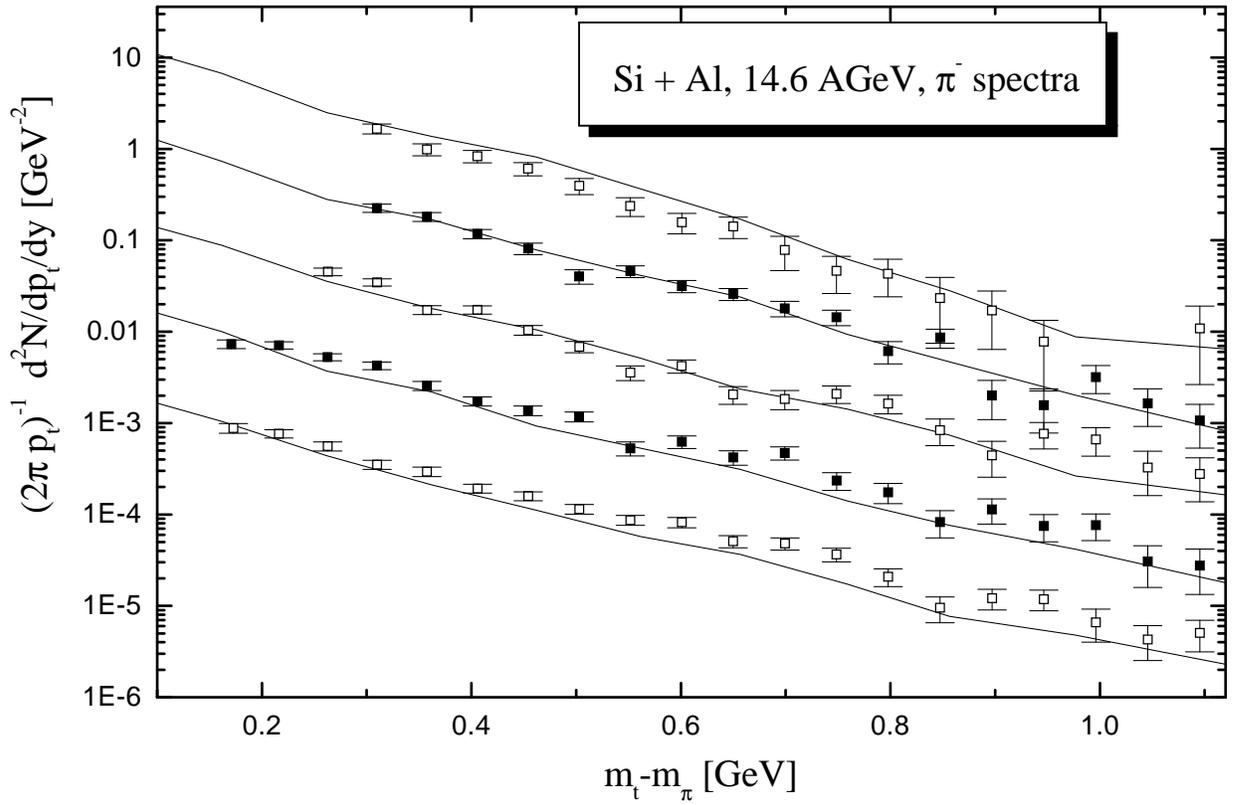,height=13cm}}
\caption{The transverse mass spectra ${1 \over 2\pi p_t}{d^2N\over dp_t dy}$ of $\pi^-$ versus
$m_t-m_\pi$ for Si~+~Al collisions at 14.6 AGeV. Results are shown for 
rapidities $y_{lab}=0.5$, 0.7, 0.9, 1.1, 1.3 and 1.5 (from top 
to bottom) successively 
scaled down by a factor of 10.
The experimental data are from Ref. \cite{abbott}.}
\label{sial-pt}
\end{figure}

\begin{figure}[ht]
\hspace{-2cm}
\centerline{\psfig{figure=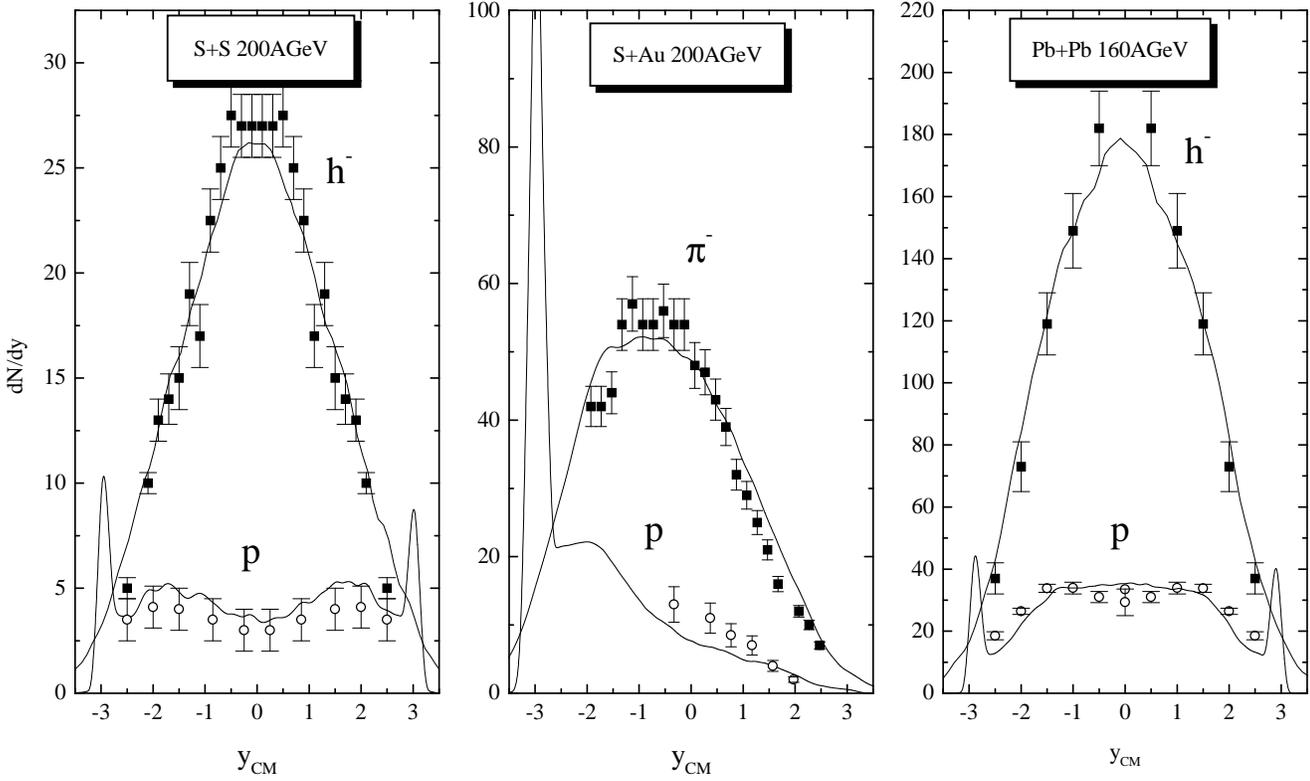,height=13cm}}
\caption{Inclusive proton and $h^-$ (or $\pi^-$) rapidity 
spectra for central S~+~S 
collisions ($b \le 1.5$ fm)
at 200 A$\cdot$GeV (left), central S~+~Au collision ($b\le 2$ fm) at
200 A$\cdot$GeV (middle) and central Pb~+~Pb collisions 
($b \le 2.5$ fm) at 160 A$\cdot$GeV (right) in comparison to the data 
from the NA35 Collaboration \cite{NA35} 
 and the NA49 Collaboration \cite{NA49}.}
\label{AB-sps-stopping}
\end{figure}

\begin{figure}[ht]
\centerline{\psfig{figure=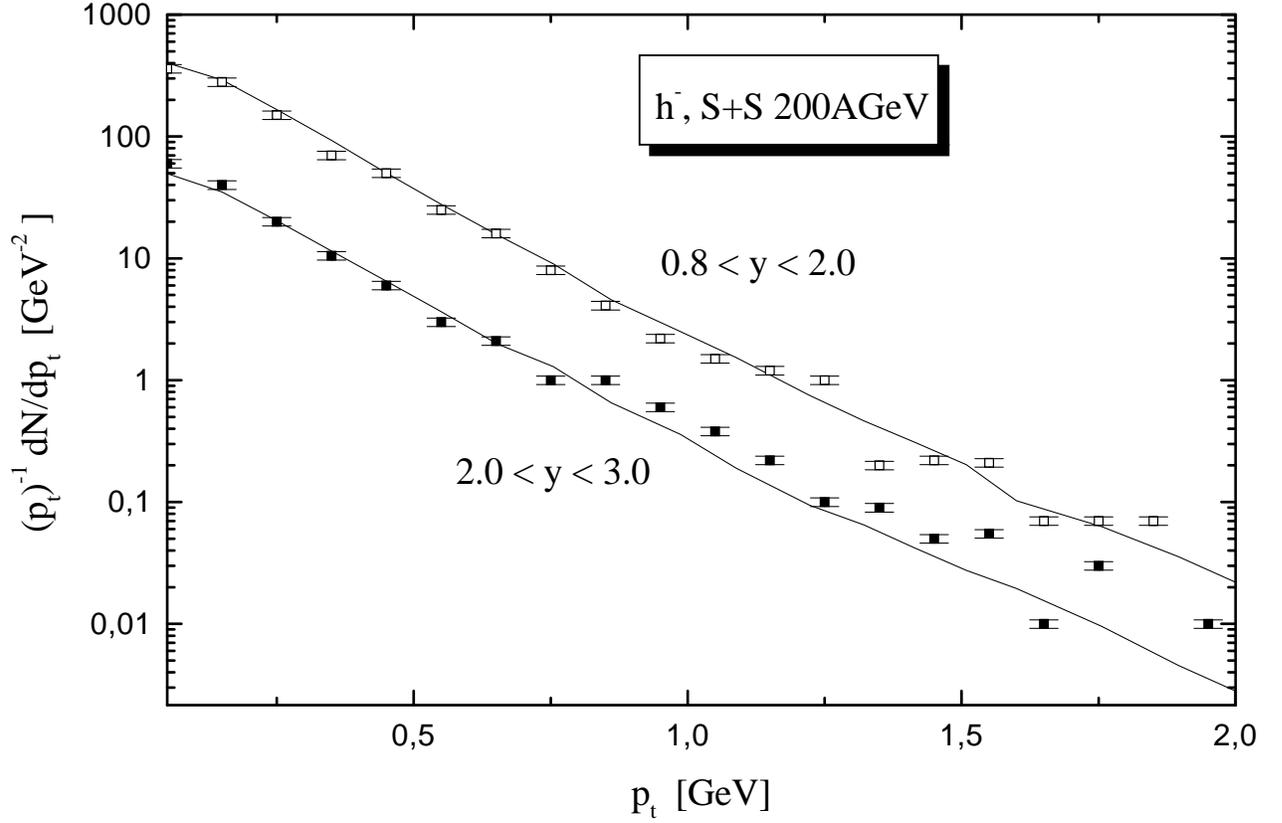,height=13cm}}
\caption{The transverse mass spectra ${1\over (p_t)}{dN\over dp_t}$ of $h^-$ versus
$m_t-m_{\pi}$ for S~+~S collisions at 200 AGeV in comparison to the
experimental data from \cite{NA35}. Results are shown for two cm rapidities
$0.8 \le y \le 2.0$ and $2.0 \le y \le 3.0$ (lower line, 
multiplied by $\times 0.1$).}
\label{ss-pt-hm}
\end{figure}

\begin{figure}[ht]
\centerline{\psfig{figure=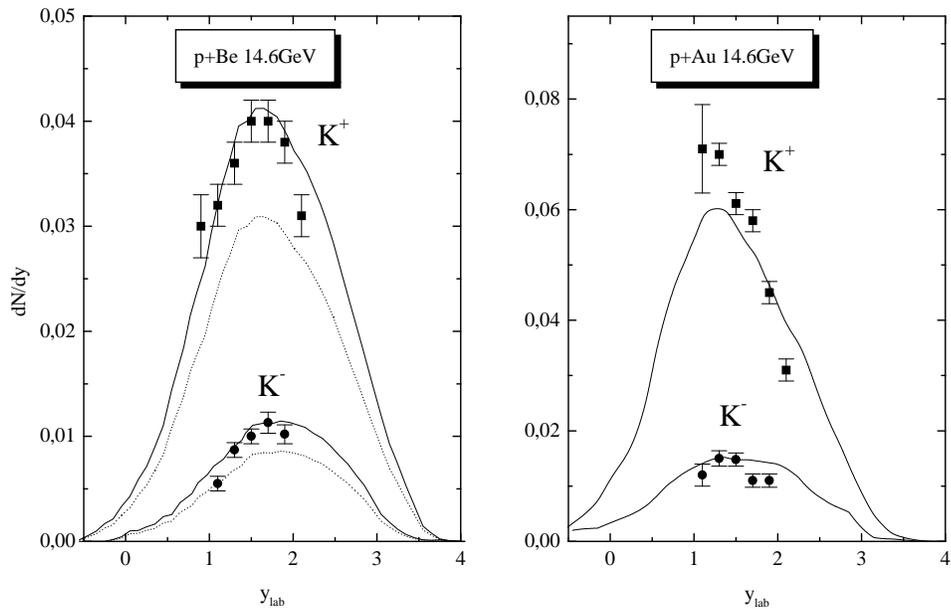,height=10cm}}
\caption{Calculated $K^+$ and $K^-$ rapidity spectra for p~+~Be (l.h.s) 
and p~+~Au (r.h.s.) at 14.6 A$\cdot$GeV (solid lines) in comparison 
to the data from the E802 Collaboration \cite{E802p}. The dashed lines for
p~+~Be are calculated with a strangeness suppression factor $\gamma_s=0.3$ 
while the solid lines are obtained for $\gamma_s =$ 0.4 (see text).}
\label{pbepau-strange}
\end{figure}

\begin{figure}[ht]
\centerline{\psfig{figure=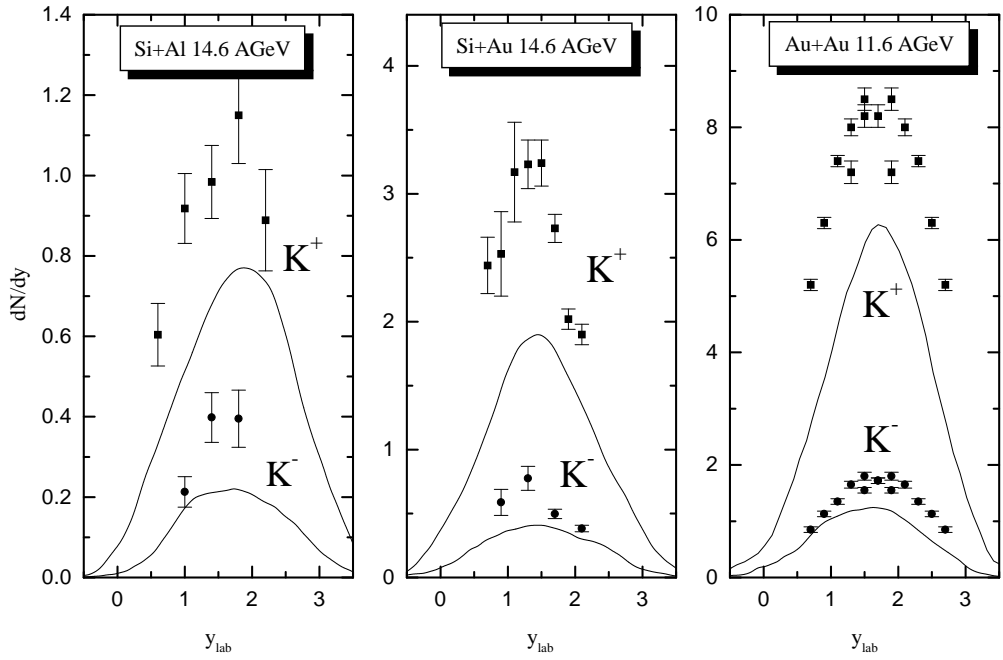,height=10cm}}
\caption{Inclusive $K^+$ and $K^-$ rapidity spectra for Si~+~Al 
at 14.6 A$\cdot$GeV (left), Si~+~Au at 14.6 A$\cdot$GeV (middle) 
and Au~+~Au at 11.6 A$\cdot$GeV (right)
in comparison to the data from the E802 Collaboration \cite{abbott} (for Si~+~Al 
and Si~+~Au) and from  \cite{Ogilvi}.}
\label{ags-strange}
\end{figure}

\begin{figure}[ht]
\centerline{\psfig{figure=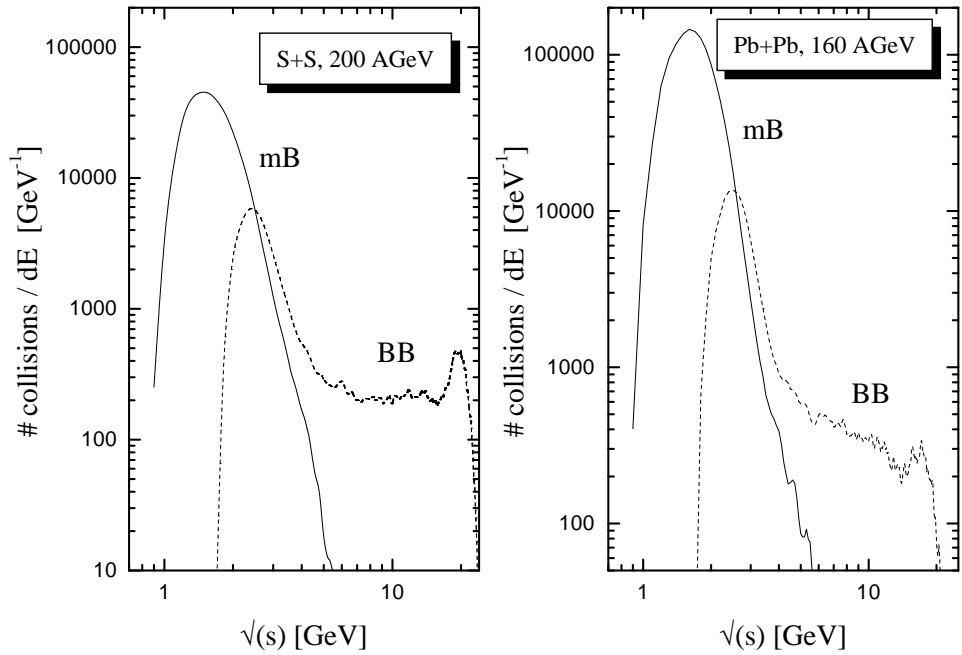,height=10cm}}
\caption{The number of baryon-baryon (dashed lines) and meson-baryon
(solid lines) 
 collisions as a function of the invariant collision energy
$\sqrt{s}$ for central S~+~S collisions (l.h.s.) 
at 200 A$\cdot$GeV and Pb~+~Pb 
collisions (r.h.s.) at 160 A$\cdot$GeV.}
\label{bb-mb}
\end{figure}

\begin{figure}[ht]
\centerline{\psfig{figure=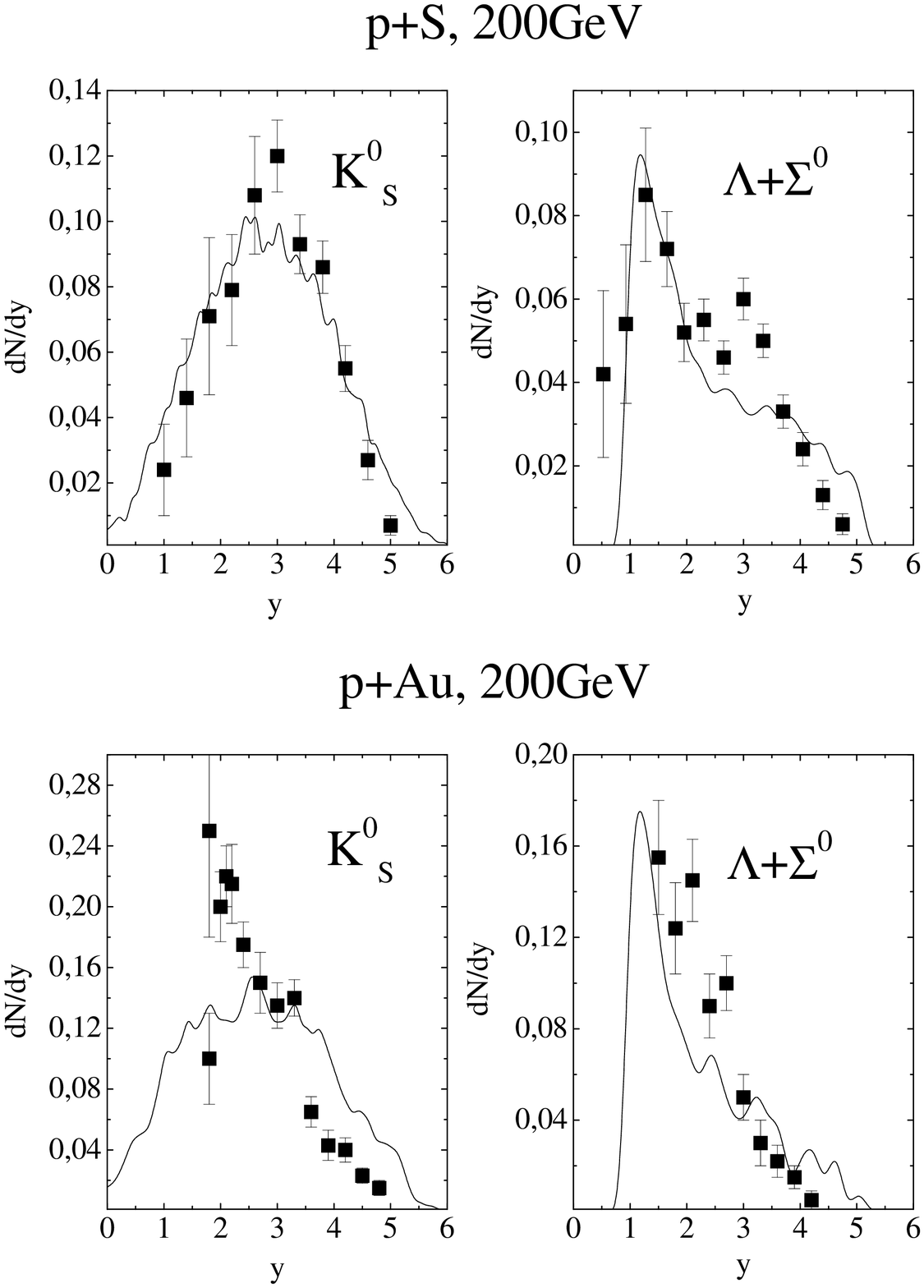,height=18cm}}
\caption{Comparison between the HSD calculations (lines) and data
from NA35 \cite{NA35pA} 
(squares) for the rapidity distribution of $K_S^0$ (left) and 
hyperons $\Lambda+\Sigma^0$ (right) produced in p~+~S collisions 
(trigger condition: more than five charged particles in the NA35
streamer chamber) and minimum bias p~+~Au collisions at 200 GeV.}
\label{pA-SPS}
\end{figure}

\begin{figure}[ht]
\centerline{\psfig{figure=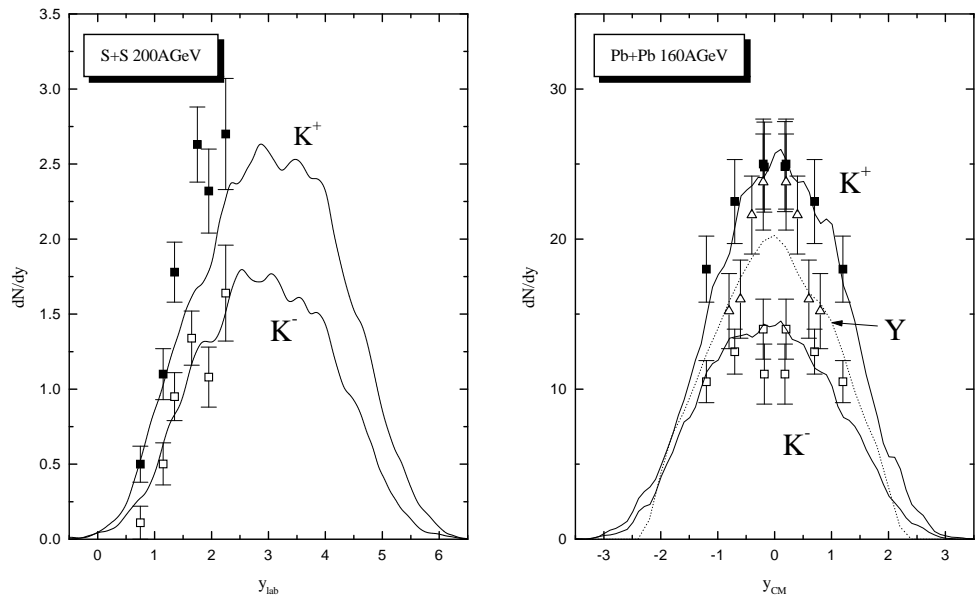,height=10cm}}
\caption{Inclusive $K^+$ and $K^-$ rapidity spectra for central S~+~S 
collisions ($b \le 1.5$ fm)
at 200 A$\cdot$GeV (left) and central Pb~+~Pb collisions 
($b \le 2.5$ fm) at 160 A$\cdot$GeV (right) in comparison to the data 
from the NA35 Collaboration \cite{NA35} 
 and from the NA49 Collaboration \cite{Bormann}. The hyperon ($Y=\Lambda,\Sigma^0$) 
rapidity 
distribution for Pb~+~Pb is shown by the dotted line in comparison to
the experimental data (open triangles) from \cite{Bormann}.}
\label{sps-strange}
\end{figure}

\begin{figure}[ht]
\centerline{\psfig{figure=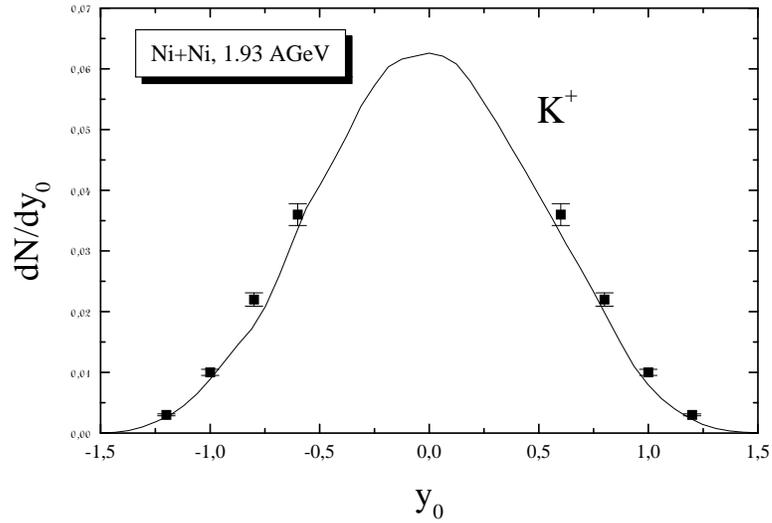,height=8cm}}
\caption{The calculated $K^+$ rapidity distribution for
Ni+Ni collisions at 1.93 AGeV as a function of
the normalized rapidity $y_0=y_{cm}/y_{proj}$ in comparison
with the FOPI data \cite{FOPI}, which have been reflected around
midrapidity.}
\label{nini}
\end{figure}

\begin{figure}[ht]
\centerline{\psfig{figure=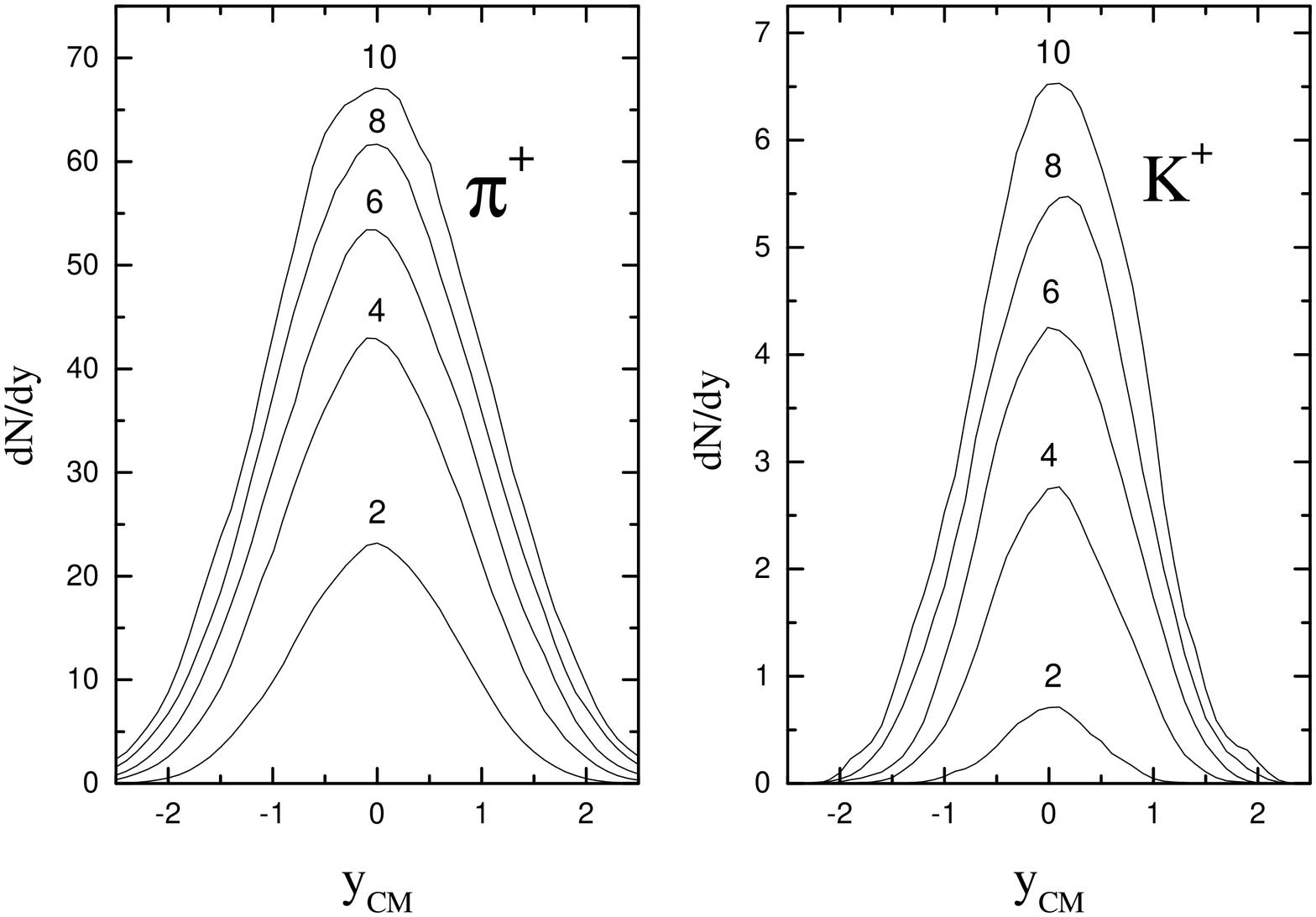,height=10cm}}
\caption{The calculated $\pi^+$ and $K^+$ rapidity distributions for
central (b=2 fm) Au~+~Au collisions at 2,4,6,8 and 10 AGeV versus
the rapidity in the cms.}
\label{auau-excitation}
\end{figure}

Furthermore, the shape of the experimental pion rapidity
distributions is narrower for A~+~A collisions compared to
p~+~p collisions, which is also not described by the hadronic
transport model. This result again  is in agreement with
findings in the RQMD model (c.f. Fig. 6 of Ref.~\cite{Gonin}) 
and with the ARC model \cite{ARC}.
Both observables, the strong
enhancement of the scaled kaon yield as well as the
lowering of the width of the pion rapidity 
distribution, are already
found for the rather light system Si~+~Al (see Fig.~\ref{AB-ags-stopping}
and Fig.~\ref{ags-strange}), where the amount of hadronic rescattering 
is rather small. Thus the shape of 
the pion rapidity distribution in Si~+~Al reactions  and the strangeness yield 
is dominated by the primary string fragmentation process,
which on the other hand is fixed by p~+~p and p~+~Be data.
The difference of p~+~Be and Si~+~Al collisions  
is hard to understand within a hadronic rescattering scenario and  
might indicate new physics for the primary $s\bar{s}$ production mechanism
at AGS energies.
 
On the other hand, the production of pions, kaons and antikaons
as well as the stopping of the incident protons at SPS energies
is in line with the hadronic transport approach. To illustrate
this we summarize in Table~\ref{tab3} the calculated kaon, antikaon
and pion yield and the scaled kaon yield 
\begin{eqnarray}
{<\!K\!> \over <\!\pi\!>}={\!<K^+ + K^- + K^0 + \bar{K^0}\! >\over 
                           <\!\pi^+ +\pi^- + \pi^0\! >}
\end{eqnarray}
in comparison with the experimental ratios from Ref.~\cite{Bormann}. 
The last two columns show a good agreement between the HSD results
and the data. Thus the signal of 'strangeness enhancement' in
A~+~A collisions at SPS energies does not qualify as a sensitive
observable for an intermediate QGP phase. Nevertheless, the
experimentally observed strong increase of the antihyperon and 
multistrange baryon yield \cite{WA94} in heavy-ion collision cannot be described
by our hadronic model, since no string fusion (as in the RQMD
version of Ref.~\cite{Hsorge}) is included to enhance this yield. 
Furthermore, only
the annihilation of antihyperons is include so far and not the inverse
production channels like e.g. $\rho+K+\bar{K} \to \Lambda +\bar{\Lambda}$
due to technical reasons. The aim of our work
was to present a systematic analysis of strangeness production 
(mainly kaons) over a wide range in energy and system mass $A_1 + A_2$. 

In order to discuss the strangeness production over the complete
energy range we also show in Table~\ref{tab3} the calculated 
$K^+/\pi^+$ ratio, which experimentally is substantially lower at SPS energies
($\approx 13.5\%$) compared to AGS energies ($\approx 19\%$)
for the most heavy systems.
At SPS energies this ratio is only enhanced by a factor 1.75 
for central Pb~+~Pb collisions compared to p~+~p reactions
and should be compared to the factor $\approx$ 3 at AGS.
Such a decrease of the scaled
kaon yield from AGS to SPS energies is hard to obtain
in a hadronic transport model. On the contrary, the higher 
temperatures and particle densities at SPS energies allways tend 
to enhance the $K^+/\pi^+$ yield closer to its thermal
equlibrium value of $\approx 20-25\%$ \cite{BraunMu,greiner} at chemical
freezeout and temperatures around $T\approx 150$ MeV.  

Our findings have to be
compared to results obtained by other microscopic approaches.
At AGS energies there exist calculations from RQMD within different 
versions \cite{Hsorge2,Gonin}
and from ARC \cite{ARC} for the system Si~+~Au. The earlier RQMD and ARC models
agree with the presented results concerning the overestimation
of the pion yield. In the kaon production there is a 
clear difference: While RQMD  \cite{Gonin} also underpredicts the kaon yield
by roughly 22\% (HSD by 30\%) the ARC code describes the kaon 
data rather well. Unfortunately the ARC code is fine tuned to heavy-ion
collisions at AGS energies. It would be interesting to investigate whether
a reduction of the scaled kaon yield at higher energies is also
described within this approach.
At SPS energies the investigations of strangeness production 
within standard RQMD and VENUS give similar results as
the HSD approach. In order to account for the observed and large 
strange antibaryons abundancies 
and multi-strange baryon abundancies \cite{WA94} conceptual new steps like  
string fusion were introduced 
as mentioned above. Unfortunately, no systematic analysis 
over the complete energy range exists from the other models. It would be of 
particular interest to compare a systematic study of  
the scaled kaon yield calculated by independent approaches to 
achieve a model independent conclusion
on the topic of strangeness production. 
 
Our systematic study of kaon production from SIS to SPS energies within a
hadronic model shows a continuous increase of the $K^+/\pi^+$ ratio 
($\approx$ 3.3\% at SIS, $\approx$ 9.5\% at AGS, $\approx$ 14\% at SPS)
with energy because the average kinetic energies and particle
densities rise with incident energy and enhance the scaled
strangeness yield. On the other hand the
energy dependence of the experimentally observed 
$K^+/\pi^+$ ratio in the most heavy systems rises from
$\approx$ 3.3\% at SIS energies to $\approx$ 19\% at AGS and drops to 
$\approx$ 14\% at SPS.
The high value observed at AGS might indicate the presence of a nonhadronic phase, 
which seems to be close to chemical equlibrium for strangeness. 
On the other side, strange antibaryon 
enhancement at SPS energies might indicate nonhadronic effects
also at SPS energies, but the small $K/\pi$ ratio (in comparison
to AGS) shows that the system is not really in full chemical 
equlibrium for strangeness.
In this context the study of strange antibaryons at AGS energies
is of particular interest. The E859 Collaboration has measured the
$\bar{\Lambda}/\bar{p}$ ratio in Si~+~Al at 14.6 A$\cdot$GeV and has recently
reported a large value
$\bar{\Lambda}/\bar{p}=2.9 \pm 0.9 \pm 0.5$ for $1.15 < y < 1.55$
\cite{E589}, which would indeed favor a scenario of (nearly) chemically saturated
strange antibaryon populations at freezeout. This ratio, however, cannot
be reached by far within hadronic cascade-type models.
\renewcommand{\arraystretch}{1.4}
\begin{table}
\centerline{\begin{tabular}{|c|c|c|c|}
\hline
\multicolumn{4}{|c|}{\rule[-3mm]{0mm}{8mm} 
\bf Au~+~Au collisions at AGS for different energies} \\ \hline 
\multicolumn{4}{|c|}{\rule[-2mm]{0cm}{6mm}
$<\!K^+\!>\! / \!<\!\pi^+\!> [\%]$} \\ \hline
$E_{lab}$ [AGeV] & HSD, full rapidity & HSD, ${|y-y_{CM}|\over y_{CM}} < 0.25$ & 
preliminary data \cite{Ogilvi}  \\ \hline\hline
2   &    3.3  & 3.6 &   4 $\pm$ 1  \\ \hline
4   &    5.6  & 5.9 &  11 $\pm$ 1  \\ \hline
6   &    7.6  & 8.0 &  14.5$\pm$ 1.5  \\ \hline
8   &    8.7  & 9.1 &  17 $\pm$ 1  \\ \hline
11  &    9.0  & 9.7 &  19 $\pm$ 1  \\ \hline
\end{tabular}}
\caption{The $K^+ / \pi^+$ yield for Au~+~Au at different incident energies 
obtained within
the HSD approach in comparison to the experimental data from Ref.~\cite{Ogilvi}.} 
\label{tab4}
\end{table}

\begin{table}
\centerline{\begin{tabular}{|c|c|c|c|l|}
\hline
\multicolumn{5}{|c|}{\rule[-3mm]{0mm}{8mm} \bf Strangeness at AGS} \\ \hline
\rule[-5mm]{0cm}{1.3cm}system& $<\!K^+\!>$ & $<\!\pi^+\!>$ 
&  $\displaystyle{ {<\!K^+\!>\! \over \!<\!\pi^+\!>} }$ HSD & 
$\displaystyle{ {<\!K^+\!>\! \over \!<\!\pi^+\!>} }$ data  \\
\hline\hline
p~+~Be  & 0.075& 1.27 & 0.059 & 0.059 $\pm$ 0.01 \\
Si~+~Al & 1.7 & 24 & 0.071 & 0.12 $\pm$ 0.01 \cite{abbott} \\
Si~+~Au & 5.2 & 63 & 0.084 & 0.17  $\pm$ 0.02 \cite{abbott} \\
Au~+~Au & 17.5 & 194 & 0.095 & 0.18  $\pm$ 0.01 \cite{Ogilvi} \\
\hline
\end{tabular}}
\caption{The $K^+ / \pi^+$ yield for different systems at AGS energies 
obtained within
the HSD approach in comparison to the experimental data. The experimental
ratio for p~+~Be is estimated by the E802 data \cite{E802p}. }
\label{tab2}
\end{table}

\begin{table}
\centerline{\begin{tabular}{|c|c|c|c|c|c|l|}
\hline
\multicolumn{7}{|c|}{\rule[-3mm]{0mm}{8mm}\bf Strangeness at SPS} \\ \hline
\rule[-5mm]{0cm}{1.3cm}system& $\!<\!K\!>\!$ & $\!<\bar{K}>\!$ 
& $\!<\pi>\! $
& $\displaystyle{\! {<\!K^+\!>\! \over \!<\!\pi^+\!>}\! }$ HSD 
& $\displaystyle{\! {<\!K\!>\! \over \!<\!\pi\!>}\! }$   HSD & 
 $\displaystyle{ {<\!K\!>\! \over \!<\!\pi\!>} }$  data  \\
\hline\hline
p~+~p  & 0.42  & 0.27 & 9.1 & 0.08  &0.08  & 0.08 $\pm$ 0.02  \\
S~+~S  & 19.65 & 12.7 & 265 & 0.11  &0.139 & 0.15 $\pm$ 0.015 \\
S~+~Au & 53    & 33.2 & 678 & 0.118 &0.13  & 0.13 $\pm$ 0.015 (for S~+~Ag)\\
Pb~+~Pb& 194   & 115  &2065 & 0.138 &0.15  & 0.14 $\pm$ 0.02 \\
\hline
\end{tabular}}
\caption{The kaon $<\! K \!>=< \! K^+ \! + \! K^0 \!>$, 
antikaon $<\! \bar{K}\! >=< \!K^- \! + \! \bar{K^0} \!>$  
and pion $<\! \pi\! >=< \!\pi^+ \! + \! \pi^- \! + \! \pi^0 \!> $ 
yield at SPS energies and the $<\! K^+ \! >/<\! \pi^+\! >$ and 
$<\! K\! >/<\! \pi\! >=
<K^+ + K^- + K^0 + \bar{K^0} >/ <\pi^+ +\pi^- + \pi^0 >$ ratio obtained by the 
HSD approach  compared to the corresponding experimental ratio taken from 
Ref.~\cite{Bormann}.}
\label{tab3}
\end{table}
\renewcommand{\arraystretch}{1.0}

\newpage

\clearpage

\end{document}